# The Cosmic Background Radiation

G.F. Smoot

*Lawrence Berkeley Laboratory and Department of Physics, University of California, Berkeley*

Observations of the Cosmic Microwave background have provided many of the most powerful constraints we have on cosmology and events in the early universe. The spectrum and isotropy of CBR have long been a pillar of Big Bang models. The discovery of low levels on anisotropy has provided new information and tools for our understanding of the early universe. Further observations promise to enhance greatly our knowledge of processes in the early universe and cosmological parameters. We can anticipate rapid advance in this field up to and through the year 2000 which will dramatically focus our efforts in cosmology during the next millenium. This paper outlines the primary science likely to be discovered and defined by a vigorous airborne and ground-based program which should be strongly supported. If successfully excuted, we an anticipate a measurement of the CBR anisotropy spectrum to within a factor of two of the confidence level unavoidably set by cosmic variance. Even so, observations of the CBR are the best and often the only way to obtain information on many critical parameters so that an ambitious satellite experiment that maps the full-sky to the $\delta T/T \sim 10^{-6}$ level is well justified.

## §1   Introduction

From its prediction (Gamow & Alpher & Hermann 1948) and discovery (Penzias & Wilson 1965) the cosmic background radiation (CBR) has been a cornerstone of the Big Bang cosmology and led to its widespread acceptance. In the intervening years improved observations have lead to a better understanding of its role in the early universe and the wealth of information that it carries about the early universe. With this work its importance to cosmology has grown rapidly while the technology to exploit it has kept pace. A major advancement came with the COBE satellite in the early 1990's with a very precise measurement of the spectrum (Mather et al. 1993) as being black body and the discovery of anisotropies (Smoot et al. 1992) at the $10^{-5}$ level. This discovery was followed by an explosion of activity and the reports of anisotropies by several groups and a rapid advance in theoretical work.

### 1.1   The CBR Spectrum

The spectrum and temperature of the cosmic background radiation are key issues in cosmology. The Big Bang model makes the unequivocal prediction that to high order the spectrum will be thermal and that the temperature will vary with redshift as $T(z) = (1+z)T_0$. The current COBE FIRAS observations now give the best evidence that the spectrum is blackbody and the best estimate of the temperature $T_0$. The precision of the spectral observations is within a factor of three of predicted distortions based upon known objects in the universe. Future improvements both from better data analysis and improved experiments are possible. There is significant new science possible with improved results.

There is much evidence for the remote presence of the cosmic background radiation from CN observations and the presence of anomalous formaldehyde absorption in distant galaxies and from the Sunyaev-Zel'dovich effect. The only evidence for the redshift dependence of the temperature comes from the Keck telescope observations of Songalia et al. (1994). The remote sensing of the cosmic background radiation is an actively emerging field. We can expect significant progress in the study of the SZ effect and the remote sensing of the CBR in the future.

### 1.2   Anisotropies in the CBR

Fluctuations in the temperature of the cosmic background radiation have now been detected over a wide range of angular scales and a consistent picture may be emerging. The existence and nature of these anisotropies have a profound impact on cosmology. First they define the kind of universe that we inhabit and how models can be developed as well as constrained. The observations of the CBR angular distribution are the best evidence that we have for the large scale isotropy and homogeneity of the universe. As such they imply that models of the universe can be treated as FRW (Friedmann-Roberstson-Walker) cosmologies - that is an expanding universe slightly perturbed around the Robertson-Walker metric. This simplifies calculations of the development of the universe from very early times to the present. The universe indicates no sign of shear and vorticity or other factors that would force us to more complicated cosmologies.

The natural, and commonly accepted, interpretation of the very small anisotropies discovered is that they result from primordial perturbations in the early universe. All indications are that the large-scale structure in the universe has developed by the process of gravitational instability from these small amplitude primordial perturbations in the energy density. Slightly overdense regions begin collapsing under the influence of gravity, becoming more and more overdense. Under dense regions rarify as they expand and their material flows to more dense regions. Over time the density contrast increases. The result is the formation of large scale structure such as galaxies, galaxy clusters, voids, and features such as the great wall of galaxies.

Observations of the CBR anisotropies provide information on these primordial perturbations and on smaller ($\leq 1°$) scales on the initial development of these perturbations. As a result it is possible to glean much new information about cosmological parameters and the content of the universe through the detailed study of the resulting temperature fluctuations. For example the angular scale ($\leq 1°$) is set by two things, (1) the speed of light - setting the largest region that can be affected by the motion of matter and energy, and (2) the speed of sound - setting the scale for regions in which matter can clump. The speed of sound is roughly $c/\sqrt{3}$ so that these two sizes are closely related. The age of the universe times this speed sets the size of these regions. The angle on the sky is set by the ratio of the age of the universe at the time the CBR last interacted to the time that we observe the CBR and by the geometry of the universe. Since the length sizes (or ages) are in ratio, any uncertainty in the distance scale (or Hubble constant) cancel and the angle on the sky is set strictly by the geometry of the universe. A flat universe predicts a peak in CBR fluctuations at an angular scale of about $\theta_{flat} = 0.8°$ (or $\ell \approx 220$). An open universe predicts a smaller angle while a closed universe predicts a larger angle simply as a consequence of the curving of light on its way to the observer. For a last scattering redshift of $\approx 1100$ the relation is simply $\theta_{peak} \cong \theta_{flat} \sqrt{\Omega_0}$ where we have assumed the standard relation between $\Omega$ and the curvature of the universe. This is as direct a measurement of the geometry of the universe as one can imagine. The corrections to the relationship are all weak.

The geometry of the universe is one of many parameters one could determine with precise measurements of the CBR anisotropies. One can expect to distinguish between models involving inflation and topological defects.

Perhaps the easiest way is (see e.g. Albrecht et al. 1995, Crittendon & Turok 1995) by comparing the existence and location of the higher order peaks in the power spectrum. However, there are other possible tests. It is also possible to distinguish among various models of inflation. Precise observations of the CBR anisotropies can in fact provide us with the opportunity to determine the parameters of inflationary models including measuring the ratio of scalar (density) to tensor (gravitational waves) perturbations, the slope of the inflaton potential during the epoch of astronomical importance, as well as the energy scale of inflation during this epoch. It is possible that these observations would provide definite evidence for the existence of gravitons. (See e.g. Knox 1995, Bond et al. 1994 and many others).

Precise observations on angular scales $\leq 1°$ can provide information on the Hubble constant, a limit on $\Omega_{baryon}$ that is comparable to and independent of the Big Bang nucleosynthesis results, information on the nature of the dark matter and evidence that the gravitational instability mechanism is working as predicted. There are many theoretical discussions smf trbored of these points (see e.g. Scott, Silk, & White 1995).

### 1.3 Polarization

The polarization of the CBR can also provide useful information about the early universe. If our current models are correct, then the level of polarization expected is discouragingly low. Existing results are at a level that indicate that there is no great inconsistency in our current models but the limits are still more than one to two orders of magnitude greater than model predictions. At the predicted level experiments are likely to be limited by instrument noise and observation time for decades to come. However, at high frequencies it is possible that the foreground will be sufficiently low as to allow discovery level observations. The polarization is likely to continue to occupy a very distant back seat to anisotropy observations.

## §2 Current Status & Anticipated Results by 2000

The standard hot Big Bang model of cosmology is supported by three observational results: the Hubble flow of distant galaxies, the abundances of the light elements, and the existence of an isotropic thermal radiation bath. But the hot Big Bang model is incomplete; the universe today is far more complex than a homogeneous soup of hydrogen, helium, neutrinos, and microwave photons. The observable matter is organized in a hierarchy of dense clumps surrounded by empty voids. Stars group into galaxies, which in turn form groups, clusters, and still larger struc-tures. Explaining our increasingly detailed observations of the large scale structure of the universe is a fundamental problem in cosmology: what is the origin of observed structures and how did they form?

To search for the origins of structure we must turn to relics from that era. The cosmic microwave background (CMB) is one such relic; its photons, through their spatial and frequency distribution, reflect the distribution of mass and energy in the early universe and record subsequent interactions between the evolving matter and radiation fields. The paradigm for structure formation consists of the gravitational infall and collapse of small "seed" perturbations in the density of the early universe. The central result of the *Cosmic Background Explorer (COBE)* was the support of this basic picture through the detection of CMB anisotropy without a corresponding distortion from a blackbody spectrum. Within this broad outline, however, a number of more detailed questions remain unanswered. No cosmological models are able to reconcile in detail the *COBE* measurements of the CMB spectrum and anisotropy with measurements of galaxy counts and peculiar velocities. What is missing from our understanding? Evidence indicates that the dominant component of the universe is "dark" matter, inferred only through its gravitational effects. What is the nature of this dark matter? Interpretation of the CMB anisotropy on co-moving scales directly comparable to the largest structures observed in optical surveys depends on the poorly constrained ionization history of the universe. In standard models the universe became neutral at redshift $z \sim 1100$ when the temperature fell below the ionization potential of hydrogen, yet the local universe is thought to be highly ionized to redshift $z \sim 5$. What is the thermal history of the universe? When did the universe become re-ionized?

## 2.1 The CBR Spectrum - Science & Status

The observed Hubble recession of galaxies has a natural interpretation as evidence for an expanding universe: in the remote past, the universe was much smaller and denser than today. At the high energies corresponding to very early times, photon-creating processes ($\gamma + e \rightarrow \gamma + \gamma + e$, $e + p \rightarrow e + p + \gamma$, $e^+ e^- \rightarrow \gamma\gamma$) proceed rapidly with respect to the Hubble expansion, creating a system in local thermodynamic equilibrium. A system in thermal equilibrium is completely characterized by its temperature and any conserved quantum numbers: in the absence of non-equilibrium interactions, the CMB will be an isotropic blackbody radiation field. In an adiabatic expansion, the photon occupation number is constant; hence, the expansion of the universe does not by itself distort the CMB spectrum, but merely scales the thermodynamic temperature $T = T_0(1+z)$.

Distortions in the CMB spectrum arise from non-equilibrium interactions of the matter and radiation fields in the evolving universe. Many such interactions are known to exist. Figure 1 illustrates an idealized thermal history of the universe. As the CMB temperature falls below the rest mass, particles pair-created at higher temperatures fall out of equilibrium and either decay, annihilate, or "freeze out" at a relic density. The resulting energy releases might be expected to leave distinctive signatures in the CMB spectrum. Unfortunately for high-energy physics, the earliest universe is unobservable to us in photons. Photon-creating processes ($\gamma + e \rightarrow \gamma' + \gamma' + e'$) proceed rapidly for times $t < 1$ year, re-thermalizing an arbitrary distortion to a new (albeit hotter) Planck spectrum. Processes at epochs with $kT_{CMB} > 2$ keV leave no significant direct signature on the CMB spectrum, and must be studied through any long-lived relics (gravitational potential variations, exotic particle species, topological defects) they might spawn.

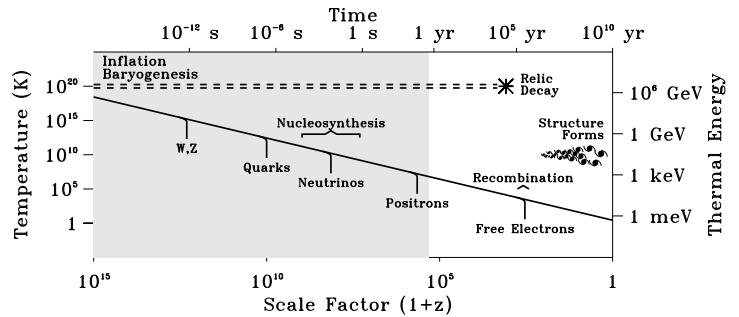

Figure 1: Idealized thermal history of the universe. The epochs at which various particle species annihilate, decay, or decouple are indicated. Events in the shaded region leave no direct signature in the CMB spectrum, but affect it though the decay of long-lived relics.

The presence today of ordered structures (galaxies, galaxy clusters, and larger-scale structure) in the face of initial isotropy is a powerful argument for such relics. The *COBE* DMR instrument shows anisotropy in the early universe ($z \sim 1100$) to be of order $\Delta T/T \sim 10^{-5}$ (Smoot et al. 1992) compared to the current clustering of order unity. Processes capable of generating structure quickly enough, without violating observational limits on the CMB isotropy, will in general release energy to the matter or radiation fields, which in turn generate distortions in the CMB spectrum. One such mechanism is "dark" non-baryonic matter, whose dynamical properties allow it to clump much more rapidly than the baryons, which only later fall into the dark-matter gravitational potential wells. Such non-baryonic dark matter may be the lightest stable member of a family of particles (e.g. su-persymmetric partners to the known particle families), in which case either the decay of the heavier unstable members or rare decay modes of the stable members can distort the CMB spectrum. Broad classes of dark-matter candidates include such decay modes. Other non-equilibrium energy-releasing processes in the early universe include the dissipation of shock and sound waves associated with primordial density or entropy perturbations, the dissipation of primeval turbulence, dissipation of gravitational wave energy, phase transitions in the early universe and associated topological defects, or energy released through isotropization of an anisotropic universe. *Any* transfer of kinetic or thermal energy between the matter and radiation fields at $t > 1$ year must alter the CMB spectrum from a blackbody distribution. The size and shape of the present distortions depend on size, redshift, and mechanism of the energy transfer, allowing observers to distinguish between various physical processes.

The most general form of CMB spectral distortion results from interactions with a hot electron gas at temperature $T_e$. Three classes of spectral distortions are particularly important, resulting from processes at different epochs. The simplest distortion is photon production from electron-ion interactions (free-free emission or thermal bremsstrahlung): $e + Z \rightarrow e + Z + \gamma$. The distortion to the present-day CMB spectrum is given by

$$\Delta T_{\mathrm{ff}} = T_\gamma \frac{Y_{\mathrm{ff}}}{x^2} \qquad (1)$$

where $T_\gamma$ is the undistorted photon temperature, $x$ is the dimensionless frequency $h\nu/kT_\gamma$, $Y_{\mathrm{ff}}$ is the optical depth to free-free emission

$$Y_{\mathrm{ff}} = \int_0^z \frac{k[T_e(z) - T_\gamma(z)]}{T_e(z)} \frac{8\pi e^6 h^2 n_e^2 g}{3m_e(kT_\gamma)^3 \sqrt{6\pi m_e kT_e}} \frac{dt}{dz'} dz', \qquad (2)$$

$n_e$ is the electron density, and $g$ is the Gaunt factor (Bartlett & Stebbins 1991). The distorted CMB spectrum is characterized by a quadratic rise in temperature at long wavelengths as the photon distribution thermalizes to the plasma temperature, and is the dominant signature for a warm plasma ($T_e \sim 10^4$ K) at recent epochs ($z<1100$).

Compton scattering ($\gamma + e \rightarrow \gamma' + e'$) of the CMB photons from the hot electron gas also contributes to CMB spectral distortions. Compton scattering transfers energy from the electrons to the photons while keeping the photon number fixed. For recent energy releases ($z<10^5$), the gas is optically thin, resulting in a CMB distortion

$$\Delta T_{\mathrm{RJ}} = T_\gamma (1 - 2y) \qquad (3)$$

in the Rayleigh-Jeans part of the spectrum where there are now too few photons, and an exponential rise in temperature in the Wien region where there are now too many

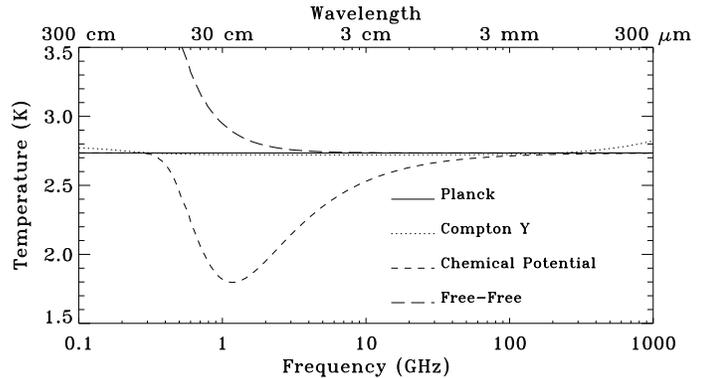

Figure 2: CMB distortions resulting from energy-releasing processes at different epochs.

photons. The magnitude of the distortion is related to the total energy transfer

$$\frac{\Delta \mathrm{E}}{\mathrm{E}} = e^{4y} - 1 \approx 4y \qquad (4)$$

(Sunyaev & Zel'dovich 1970), where the parameter $y$ is given by the integral

$$y = \int_0^z \frac{k[T_e(z) - T_\gamma(z)]}{m_e c^2} \sigma_T n_e(z) c \frac{dt}{dz'} dz', \qquad (5)$$

of the electron pressure $n_e kT_e$ along the line of sight and $\sigma_T$ denotes the Thomson cross section (Zel'dovich & Sunyaev 1969).

Compton scattering alters the photon energy distribution while conserving photon number. After many scatterings the system will reach statistical (not thermodynamic) equilibrium, described by the Bose-Einstein distribution

$$\eta = \frac{1}{e^{x+\mu_0} - 1} \qquad (6)$$

with dimensionless chemical potential

$$\mu_0 = 1.4 \frac{\Delta \mathrm{E}}{\mathrm{E}}. \qquad (7)$$

Free-free emission thermalizes the spectrum to the plasma temperature at long wavelengths. Including this effect, the chemical potential becomes frequency-dependent,

$$\mu(x) = \mu_0 \exp\left(-\frac{2x_b}{x}\right), \qquad (8)$$

where $x_b$ is the transition frequency at which Compton scattering of photons to higher frequencies is balanced by free-free creation of new photons. The resulting spectrum has a sharp drop in brightness temperature at centimeter wavelengths (Burigana et al. 1991).

The different distributions probe energy releases at different epochs, and hence different physical processes.

Figure 2 shows the distortion to the CMB spectrum corresponding to each distribution. The equilibrium Bose-Einstein distribution results from the oldest non-equilibrium processes ($10^5 < z < 8 \times 10^6$) such as the decay of relic particles or primordial anisotropy. Energy releases at more recent epochs ($z < 10^5$) produce a Comptonized spectrum, a prime candidate being a hot ($T_e > 10^5$ K) intergalactic medium. Free-free emission dominates for recent reionization ($z < 10^3$) from a warm IGM. The different physical processes distort the CMB spectrum at different wavelength regimes. Measurements near the CMB peak at millimeter wavelengths are sensitive primarily to the Compton processes $y$ and $\mu$. Measurements at longer wavelengths are sensitive to plasma bremsstrahlung from a recent reionization or to the earliest energy releases from relic decay. Measurements at both long and short wavelengths are required to constrain fully the thermal history of the universe.

Figure 3 summarizes precise measurements of the CMB spectrum from ground-based, balloon, and space platforms. The *COBE*-FIRAS instrument provides a precise determination of the CMB spectrum at mm and sub-mm wavelengths, while ground-based and balloon-borne radiometers measure the spectrum to much lower precision at centimeter and longer wavelengths. The CMB spectrum is consistent with a blackbody across more than 3 decades of frequency: a least-squares fit to all CMB measurements yields limits

$$T_\gamma = 2.73 \pm 0.01 \text{ K}$$
$$|y| < 2.5 \times 10^{-5}$$
$$|\mu_0| < 3.3 \times 10^{-4}$$
$$|Y_{\text{ff}}| < 1.9 \times 10^{-5}$$

at 95% confidence, corresponding to limits on energetic processes $\Delta E/E < 2 \times 10^{-4}$ occurring between redshifts $10^3$ and $8 \times 10^6$ (Mather et al. 1994, Wright et al. 1994, Bersanelli et al. 1994). The early universe, apparently, was a quiet place, with most of the energy residing in an isotropic radiation bath and the dark matter. The large scale structure and voids observed in optical galaxy surveys did not result from primeval explosions, since the energy required is more than 2 orders of magnitude greater than allowed. There is no uniform, hot intergalactic medium: the 35 keV electrons that produce the diffuse X-ray background have a volume filling factor below $10^{-4}$ (Wright et al. 1994).

The *COBE* results limit deviations from a blackbody spectrum to less than 0.03% of the peak CMB intensity (less than 1 mK in temperature units) over the 0.05–5 mm wavelength range, but results at longer wavelengths are much less precise: distortions as large as several percent could exist at centimeter wavelengths or longer without violating existing data. The *COBE* limits on spectral distortions at mm wavelengths do *not* imply correspond-ingly small deviations at centimeter wavelengths. In fact, results at cm wavelengths lie systematically colder than *COBE* (Kogut et al. 1991). Although suggestive of a possible distortion, the existing long-wavelength data lack sufficient precision to discriminate between a small distortion and an error in the relative calibration between the *COBE* and ground-based results (Kogut 1992).

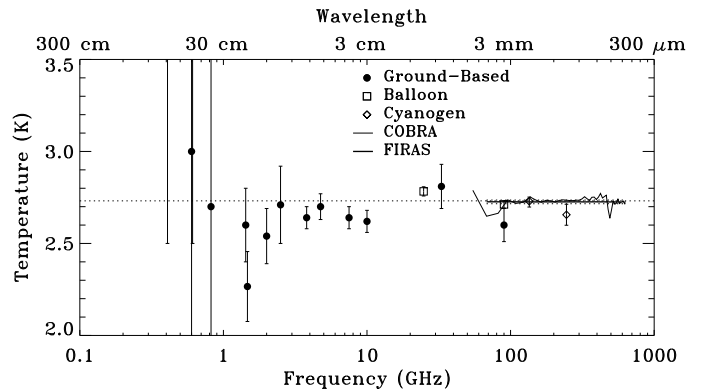

Figure 3: Precise measurements of the CMB thermodynamic temperature. The dotted line represents a 2.73 K blackbody. The CMB spectrum is poorly constrained at centimeter or longer wavelengths.

Nor have the current observational results exhausted the supply of interesting cosmological questions. Among the most urgent questions is the thermal history of the universe: when did the universe recombine? In the simplest model, recombination occurs at redshift $z \sim 1100$ when the CMB temperature falls below the ionization potential of hydrogen, leaving the bulk of the matter in a neutral state. The direct observational evidence for the standard thermal history, though, is weak. The Gunn-Peterson test shows no evidence for significant neutral hydrogen column density in the intergalactic medium, indicating that the local universe is highly ionized to $z \sim 5$. The *COBE* limit requires only that the total energy released during reionization be less than $4y$ times the energy in the CMB. Although the energy per ionization is $\sim 10$ eV, there are $10^9$ CMB photons per electron with energy $10^{-2}(1+z)$ eV. The COBE energetics requirement $\Delta E/E < 10^{-4}$ for a single reionization is easily met. Structure formation through photoionization from the first generation of collapsed objects could plausibly reionize the universe. Indeed, reionization at $z \sim 50$ is almost inevitable in cold dark matter models of structure formation and could occur as early as $z \sim 200$ in non-Gaussian isocurvature models, including topological defect models (Tegmark, Silk, & Blanchard 1994).

The ionization history of the universe depends upon the rare high-amplitude peaks in the density field, and hence provides important information for models of structure formation. If evidence for reionization is found, it probes the physics of the heating mechanism, while a negative result confirms the standard thermal history (the universe is neutral for 50<$z$<1100), ruling out many alternate cosmologies. Fixing the thermal history complements efforts to interpret the CMB anisotropy: at degree angular scales, pre-existing anisotropies will be smoothed by Compton scattering as they pass through the plasma, reducing their amplitude. The reduced amplitude from reionization can be mistaken for evidence of other physical processes, including gravity-waves, a larger Hubble constant, or an intrinsic reduction in primordial power at small angular scales (White, Krauss, & Silk 1994, Bond et al. 1994).

We may observe a recent reionization through its distinctive spectral signatures, requiring measurements of the CMB spectrum at both long and short wavelengths. The *COBE* limits to Compton-distorted spectra at millimeter wavelengths limit the plasma temperature from above, $\Delta T_y \propto n_e k T_e$. Precise new measurements at centimeter wavelengths of the free-free emission from the same reionized gas limit the plasma temperature from below, $\Delta T_{\text{ff}} \propto n_e^2/\sqrt{T_e}$. Together, the combined results can either detect the direct signature of reionization or rule out all ionized models except those with temperatures $T_e \sim T_\gamma$ (Bartlett & Stebbins 1991). Given the current *COBE* limits on $y$, a determination of the thermal history at $z \sim 50$ requires a measurement of the free-free parameter $Y_{\text{ff}}$<$10^{-7}$, corresponding to a 100 $\mu$K spectral distortion at 10 cm wavelength. This is obtainable within reasonable experimental sensitivity.

A second question of interest is the power spectrum of primordial density perturbations. The *COBE* detection of CMB anisotropy on angular scales above 7° (Smoot et al. 1992) provides important support for structure formation via gravitational instability. The *COBE* anisotropy data are well-described by a Gaussian primordial density field with power spectrum $P(k) \propto k^n$ per comoving wave number $k$, with power-law index $n = 1.0 \pm 0.4$ (Górski et al. 1994). Measurements of the CMB spectrum complement direct measurements of the density power spectrum. Short-wavelength fluctuations which enter the horizon while the universe is radiation-dominated oscillate as acoustic waves of constant amplitude and are damped by photon diffusion, transferring energy from the acoustic waves to the CMB spectrum and creating a chemical potential. The energy transferred, and hence the magnitude of the present distortion to the CMB spectrum, depends on the amplitude of the perturbations as they enter the horizon through the power-law index $n$ (Daly 1992; Hu, Scott, & Silk 1994). Although the CMB spectral distortion predicted by the inflationary value $n = 1$ is unobservably small, models with "tilted" spectra $n$>1 can produce observable distortions. Probing the range $n$<1.4 of direct interest to non-standard models requires a determination of the chemical potential $|\mu_0|$<$10^{-5}$, corresponding to a 0.5 mK spectral distortion at 10 cm wavelength.

Exotic particle decay provides another source for non-zero chemical potential. The dynamics of clusters of galaxies requires a stronger gravitational field than can be inferred by luminous matter, strongly suggesting the existence of a "dark" component of the matter field. Evidence exists that much of this dark matter is non-baryonic. Particle physics provides a number of dark matter candidates, including massive neutrinos, photinos, axions, or other weakly interacting massive particles (WIMPs). In most of these models, the current dark matter consists of the lightest stable member of a family of related particles, produced by pair creation in the early universe. Decay of the heavier, unstable members to a photon or charged particle branch will distort the CMB spectrum provided the particle lifetime is greater than a year. Rare decays of stable particles (e.g., a small branching ratio for massive neutrino decay) provide a continuous energy input and also distort the CMB spectrum. The size and wavelength of the CMB distortion are dependent upon the decay mass difference, branching ratio, and lifetime. Stringent limits on the energy released by exotic particle decay provides an important input to high-energy theories including supersymmetry and neutrino physics.

At present work in the CBR spectrum is relatively limited. The COBE team is still analyzing the FIRAS data but with a reduced staff. We can hope that a successful analysis could lead to as much as a factor of three improvement in the current COBE FIRAS limits or measurements. That is the practical limit and still leaves a significant portion of the spectrum at longer wavelengths to be probed with precision. Currently only a couple of groups are actively making spectrum observations. At the present time, it appears no groups are funded for such measurements for the next few years with the exception of internal GSFC award for the ARCADE balloon-borne instrument as a possible predecessor for a satellite. Suzanne Staggs (U. Chicago) and colleagues are planning a 10 GHz balloon measurement. At this point the CBR spectrum field has been the victim of its own success in terms of precise experiments finding thus far null results. The results have been and continue to be powerful constraints on cosmological models. The need for better measurements in the cm wavelength region and the prospects for a satellite are covered in the longer term future section.

## 2.2 Anisotropies in the CBR - Science & Status

The origin of large scale structure in the Universe is one of the key fundamental issues in cosmology. Gravitational instability models hold that large-scale structure forms as the result of gravitational amplification of initially small perturbations in the primordial density distribution. These density perturbations leave their imprint as anisotropies in the Cosmic Microwave Background (CMB) radiation. The *COBE* Differential Microwave Radiometers (DMR) maps of the cosmic microwave background anisotropy measure the primordial density distribution on super-horizon scales and offer a unique probe of the "initial conditions" for structure formation. On smaller angular scales there can be significant processing and movement of the primordial perturbations. The resulting CMB anisotropy power spectrum is rich in content and information on these processes and on cosmological parameters.

Anisotropies in the cosmic microwave background (CMB) probe the distribution of mass and energy in the early universe and provide a discriminant of competing models of structure formation. In inflationary models, the large-scale CMB anisotropy results from quantum fluctuations which are stretched in the accelerating expansion of the Universe to astronomical scales. Competing models (topological defects, axions, late-time phase transitions) generally involve relic residual high-energy vacuum state and have higher-order correlations and non-Gaussian distributions.

In general, it is thought that the cosmic background radiation photons travel to us in the present without significant interaction with matter from a redshift of about $z_{ls} = 1100$. This is the redshift for which the high energy tail of the cosmic background photons can ionize hydrogen. Above $z_{ls} = 1100$ the universe is ionized and the photons and baryon-electron plasma interact through Thomson scattering. The transition of the cosmic radiation from a collisional regime to free photons takes place in a time on the order of the Hubble time at that epoch. The last scattering region is a shell of finite thickness in redshift. If the decoupling occurred at a relatively large redshift ($z \geq 1000$), the thickness of the shell is roughly 1/15 of the mean redshift, which is relatively narrow ($\approx 10'$) from the observers point of view. Because of the strong thermal contact between the radiation, there is a compensation that makes the shell essentially equivalent to a surface and it is often treated and named as such. The last-scattering surface is generally taken to be optical depth unity for Thomson scattering. This is about equal to the visibility depth as Thomson scattering is nearly isotropic, so that a single scattering erases most previous anisotropy.

Causally-connected regions at the surface of last scattering, as viewed from the present epoch, subtend an angle $\theta \sim 1°.7\, \Omega_0^{1/2}\, (\,1100/(1+z_{ls})\,)^{1/2}$. Anisotropy on larger angular scales thus reflects primordial conditions unaffected by local physics in the photon-baryon fluid and offers a look at the primordial fluctuations. Anisotropy on smaller angular scales provides a glimpse at the processing and other effects allowing a test between competing models.

## CMB Temperature Anisotropies from Primordial Perturbations

If the cosmic background has a black-body (Planckian) distribution locally at the last scattering surface, the occupation number for each mode is

$$n(\nu) = \frac{1}{e^{h\nu/kT} - 1}. \qquad (9)$$

If the photons undergo a redshift $z$ in traveling from that location of last scattering to the receiver, then the received frequency, $\nu_R$, is related to the emitted frequency, $\nu_E$, according to:

$$\frac{\lambda_R}{\lambda_E} = \frac{\nu_E}{\nu_R} = (1+z). \qquad (10)$$

Since the number of photons per mode will be conserved,

$$\begin{aligned} n(\nu_R) &= n(\nu_E) = \frac{1}{e^{h\nu_E/kT_E} - 1} \\ &= \frac{1}{e^{h\nu_R(1+z)/kT_E} - 1} = \frac{1}{e^{h\nu_R/kT_R} - 1}, \end{aligned} \qquad (11)$$

which implies that the relation between the temperature on the emitting or last-scattering surface, $T_E$, and the temperature observed by a receiver, $T_R$, is

$$\frac{T_R}{T_E} = \frac{1}{1+z} = \frac{(k^\alpha u_\alpha)_R}{(k^\alpha u_\alpha)_E} \qquad (12)$$

where $u_\alpha$ is the four-vector velocity of the observer or emitter and $k^\alpha$ is the vector tangent to the null geodesic (photon path) connecting the events of emission and observation. The dominant effect is the cooling by redshift. To first order the variation in observed CMB temperature, $T_R$, (assuming a Planckian distribution) is set by the variation in emitting temperature and the variation in redshift according to the relation

$$\frac{\delta T_R}{T_R} \simeq \frac{\delta T_E}{T_E} - \frac{\delta z}{1+z} \qquad (13)$$

where $T_E$ is the temperature on the surface of emission (the last scattering surface in general for the CMB), and $(1+z)$ is the redshift from the observer (receiver) $R$ to the surface of emission $E$. The primary effect is the expansion of the Universe which causes both the initial $E$ temperatures and temperature variations to decrease by the factor of $(1+z)$ leaving the ratio unchanged. If the redshift to the emitting surface varies with direction, $\delta z(\theta, \phi)$, then there will be a corresponding temperature variation with direction.

The equivalence principle tells us that there will be a gravitational redshift from the matter-energy density perturbations. The Doppler effect produces a frequency shift arising from the difference in velocity between the emitter and observer. Varying velocities on the surface of last scattering will result in changes in different directions. There are thus three effects identified here as causing temperature anisotropy: (1) primordial (at last scattering surface) temperature variation (2) varying redshift with location due to gravitational potential variations, and (3) motion of the emitter and observer. The varying redshift can be caused by scalar, vector, and tensor perturbations (i.e. density variations, vorticity, and gravitational waves respectively) or general anisotropic expansion. One can just write down an equation for the received temperature as the sum of these effects. Often these effects are treated separately, but one must be careful with questions of coordinate system (choice of gauge) and treatment, particularly in the case of unusual cosmologies. A full General Relativistic treatment shows that one can trade the expansion redshift, gravitational redshift, and Doppler shift for each other by choice of coordinates but that the total result should be independent of coordinates with proper treatment. This results from the equivalence principle.

We will focus here on the anisotropies of primordial density perturbations in a near Friedmann-Robertson-Walker (FRW) universe. The high degree of CMB isotropy (e.g. Smoot et al. 1992) indicates that this is an appropriate approach (Stoeger, Maartens, & Ellis 1994). The standard practice is to use conformal time, linearize the problem in an expanding universe, and use the fact that light travels on a null geodesic. The conformal time, $\eta$ is defined as $d\eta \equiv dt/a(t)$, where $a(t)$ is the scale factor for the Universe. Then $g_{\mu\nu}$, linearly perturbed from the Robertson-Walker metric, $ds^2 = g_{\mu\nu}dx^\mu dx^\nu \cong a^2(\eta)[\eta_{\mu\nu} + h_{\mu\nu}]dx^\mu dx^\nu$, where $\eta_{\mu\nu}$ is the Minkowski metric.

Sachs and Wolfe (1967) gave the first and an excellent treatment of the CMB anisotropies that result from potential variations in a FRW space-time, though their results are some what more general. They derived the relation good to first order:

$$T_R = T_E \frac{\eta_E^2}{\eta_R^2}(1 + \frac{\delta T_R}{T_R}), \qquad (14)$$

where $\eta$ is the conformal time and the subscripts $R$ stands for received and $E$ for emitted or last scattered location epochs and

$$\frac{\delta T_R}{T_R} = \frac{1}{2}\int_0^{\eta_R-\eta_E}(\frac{\partial h_{\mu\beta}}{\partial \eta}e^\mu e^\beta - 2\frac{\partial h_{\beta 0}}{\partial \eta}e^\beta)_{(o)}dy, \qquad (15)$$

where $\delta T_R/T_R$ includes all the effects via using the null geodesic and $e^\beta$ is the direction vector of the light the receiver sees. The Sachs-Wolfe calculation is generally separated into parts: gravitational redshift from potential variation on the surface of last scattering, the time variation of potentials along the photon's path, the Doppler effect due to the relative motion of the emitter and observer (to first order in $v/c$), and variations in the temperature or number density of photons. This can be written as

$$\frac{\delta T_R}{T_R} = \frac{\delta \Phi}{c^2} + \frac{2}{c}\int_E^R d\tau \frac{\partial \Phi}{\partial \tau}(\tau, \mathbf{x}) + \mathbf{n}\cdot\frac{(\mathbf{v}_R - \mathbf{v}_E)}{c} + \frac{\delta T_E}{T_E} \qquad (16)$$

where the first term is the cosmological gravitational redshift, the second term is for the time changing gravitational potential along the photons' path, the next is the Doppler effect of the receiver and emitter relative velocities along the line of sight, and the last term is the variation in temperature with emitter location (e.g. see Appendix B of White, Scott, & Silk 1994).

### Anisotropy from Adiabatic Density Fluctuations

We now consider the effect of adiabatic density fluctuations. Adiabatic fluctuations are those in which all constituents (photons, baryons, and whatever dark matter) maintain a constant (number) density ratio. Thus the photon density, and therefore temperature, vary in tandem with the potential. Since the photon energy density $\rho_\gamma \propto T^4$, the variation in temperature is

$$\frac{\delta T_E}{T_E} = \frac{1}{4}\frac{\delta \rho_\gamma}{\rho_\gamma}|_E. \qquad (17)$$

The potential, $\Phi$, is related to the density field $\rho(x,t)$ via the Poisson equation,

$$\frac{1}{a^2}\nabla^2\Phi = 4\pi G\rho, \qquad (18)$$

where $a$ is the cosmological scale factor and $\nabla^2$ is the Laplacian with respect to comoving coordinates. In the Newtonian limit $\Phi = GM/r$ so that fluctuations in density, $\delta\rho$ result in potential fluctuations $\delta\Phi$. The relationship is readily found in the following way:

$$\delta\Phi = \frac{G\delta M}{r} = G\frac{\delta M}{M}\frac{M}{r} = G\frac{\delta\rho}{\rho}\frac{4\pi r^2}{3}\rho. \qquad (19)$$

The Hubble expansion gives us the relation between the Hubble expansion rate, $H$, and the critical density, $\rho_c$, namely $\rho_c = 3H^2/8\pi G$ which we can substitute into the equation above yielding:

$$\delta\Phi = \frac{1}{2}\frac{\delta\rho}{\rho}(Hr)^2\frac{\rho}{\rho_c} \quad or \quad \frac{\delta\rho}{\rho} = 2\delta\Phi\frac{1}{(Hr)^2}\frac{\rho_c}{\rho} \qquad (20)$$

We make the approximation that $\rho \approx \rho_c$ which is more and more accurate as one goes back in time. Even if $\Omega_o \approx 0.01$, which is the lower limit set by visible matter, then $\rho$ will be 90% of $\rho_c$ by a redshift of 1000. At scales $r > c/H$ (larger than the horizon) most of the contribution comes from distances comparable to $c/H$ because causality limits the range over which gravitation can act so that $\delta\rho/\rho \sim 2\delta\Phi/c^2$. (Note than in the weak field approximation for comoving coordinates $h_{00} = 2\delta\Phi$ which indicates the geodesic approach will get the same answer.) Thus for adiabatic perturbations $\delta\rho/\rho \approx 2\delta\Phi/c^2$ and since $\delta\rho_\gamma/\rho_\gamma = 4/3\,\delta\rho/\rho$, then $\delta\rho_\gamma/\rho_\gamma|_E = -8/3\,\delta\Phi/c^2$.

In this case the Sachs-Wolfe calculation gives

$$\frac{\delta T_R}{T_R} = \frac{1}{c^2}\delta\Phi + \frac{2}{c}\int_E^R d\tau \frac{\partial\Phi}{\partial\tau}(\tau,\mathbf{x}) + \mathbf{n}\cdot(\mathbf{v}_R - \mathbf{v}_E)/c + \frac{1}{4}\frac{\delta\rho_\gamma}{\rho_\gamma}|_E \quad (21)$$

$$\frac{\delta T_R}{T_R} = \frac{1}{3c^2}\delta\Phi + \frac{2}{c}\int_E^R d\tau \frac{\partial\Phi}{\partial\tau}(\tau,\mathbf{x}) + \mathbf{n}\cdot(\mathbf{v}_R - \mathbf{v}_E)/c \quad (22)$$

For most circumstances the Sachs-Wolfe effect is dominated by the conditions on the surface of last scattering. The effect of the time varying potentials (called the Rees-Sciama effect) on the photons is generally an order of magnitude less than the last-scattering surface effect of perturbations. When the perturbations are in the linear regime, the time rate of change of the potential is zero and there is no Rees-Sciama effect.

In this section we focus on the CMB anisotropy generated by the gravitational potential differences created by adiabatic fluctuations. We leave the treatment of relative velocities to a later section. This is generally valid for anisotropies produced by adiabatic fluctuations on scales larger than the horizon. Adiabatic fluctuation potential variations, $\delta\Phi$, on the surface of last scattering give anisotropies:

$$\frac{\delta T}{T} = \frac{1}{3}\frac{\delta\Phi}{c^2} \quad (23)$$

which are in turn related to the density field $\rho(\mathbf{x},t)$ via the Poisson equation above.

If the primordial perturbations are initially small, $\rho(\mathbf{x},t) = \overline{\rho}\,[1 + \delta(\mathbf{x},t)]$, we can linearize the treatment and Fourier expand the density perturbations $\delta(\mathbf{x})$,

$$\delta(\mathbf{x}) = \sum_k \delta_k\,e^{i\mathbf{k}\cdot\mathbf{x}\,-\,\phi_k} \rightarrow \frac{V}{(2\pi)^3}\int_{Vol} \delta_k\,e^{i\mathbf{k}\cdot\mathbf{x}\,-\,\phi_k} d^3k \quad (24)$$

where, by the Fourier transform,

$$\delta_k = V^{-1}\int_{Vol}\delta(\mathbf{x})\,e^{i\mathbf{k}\cdot\mathbf{x}}d^3x \quad (25)$$

The density field is Gaussian if the amplitude probability distribution at any spatial point follows a Gaussian distribution

$$p(\delta) = \frac{1}{\sqrt{2\pi\sigma^2}}\,\exp(-\frac{\delta^2}{2\sigma^2}) \quad (26)$$

with a uniform distribution in phase, where $\delta_k = A(k)e^{i\phi_k}$,

$$0 < \phi_k \leq 2\pi. \quad (27)$$

This is expected to be the case for most sources of perturbations and in particular for nearly all inflationary models. Gaussian fields have the desirable property that they are completely specified by the power spectrum $P(k) = \langle|\delta_k|^2\rangle$ or its Fourier transform, the 2-point correlation function. To this point all tests on the COBE DMR data show them to be consistent with Gaussian statistics (Smoot et al. 1994, Hinshaw et al. 1994, Kogut et al. 1994, Kogut 1995). The significance of these tests for all angular scales is not yet clear; however, we will assume it as a working assumption. If it is not true, then inconsistencies will appear, when new, high-quality CMB anisotropy data become available. Though the power spectrum will show anomalies, it is likely that other tests for non-Gaussian fluctuations and non-random phase will be more powerful.

We can now use the first term in the Sachs-Wolfe effect to calculate the CMB anisotropy from the power spectrum of density perturbations

$$\frac{\delta T}{T} = \frac{1}{3}\frac{\delta\Phi}{c^2} = -\frac{a_0^2 H_0^2}{2(2\pi)^3}\int k^{-2}\delta_k\,e^{i\mathbf{k}\cdot\mathbf{x}\,-\,\phi_k}d^3k, \quad (28)$$

where the vector $\mathbf{x}$ points to the last-scattering surface and has length $2cH_0^{-1}$, $H_0$ is the current Hubble expansion rate, and $a_0$ is the current scale size of the Universe.

### 2.3 Spherical Harmonic Decomposition & Power Spectrum

The CMB anisotropy may be decomposed into a spherical harmonic representation

$$T(\theta,\phi) = \sum_{\ell m} a_{\ell m} Y_{\ell m}(\theta,\phi) \quad (29)$$

where $\theta$ and $\phi$ are the spherical angles on the sky. To first order the spherical harmonic coefficients $a_{\ell m}$ are given by

$$\langle|a_{\ell m}|^2\rangle = V^{-1}\frac{H_0^4}{2\pi}\int_0^\infty \frac{dk}{k^2}|\delta_k|^2[j_l(kx)]^2, \quad (30)$$

where $j_l$ is the spherical Bessel function of order $l$, and the average is over all possible realizations of fluctuations. One expects this is equivalent to an average over all observation positions in the Universe. This is justified by the assumption that we do not occupy a special position - the Copernican Principle. This average over all realizations (observation position) results in random phases for the various components.

If we assume that the initial power spectrum is a power law:

$$|\delta_k|^2 = AVk^n, \quad (31)$$

the integral over $[j_l(kx)]^2$ yields

$$\langle a_{\ell m}^2\rangle = \frac{AH_0^{n+3}}{16}\frac{\Gamma[l+(n-1)/2]\,\Gamma[3-n]}{\Gamma[l+(5-n)/2]\,\Gamma[(4-n)/2]^2} \quad (32)$$

$$= (Q_{rms-PS})^2 \frac{4\pi}{5} \frac{\Gamma[l+(n-1)/2]\,\Gamma[(9-n)/2]}{\Gamma[l+(5-n)/2]\,\Gamma[(3+n)/2]} \quad (33)$$

(Kolb & Turner 1991, Bond & Efstathiou 1987). Up to a numerical factor $\langle a_{\ell m}^2 \rangle^{1/2}$ is equal to the value of $\delta\rho/\rho$ on the present horizon.

In this case the spherical harmonic coefficients $a_{\ell m}$ are Gaussian random variables with zero mean and $\ell$-dependent variance given by $\langle a_{\ell m}^2 \rangle$.

The power spectrum from a power law of primordial perturbations is then simply given by the above formula for perturbations larger than the horizon at last scattering. For smaller angular scales it is no longer accurate both because of the approximation used and because there is perturbation processing.

### Anisotropy from Perturbation Processing

The primary perturbation processing in the early universe is the acoustic oscillation of the photon-baryon fluid. A positive perturbation will provide an extra gravitational attraction on its surrounding material, which will cause material to flow towards the inhomogeneity. The density fluctuation will continue to pull in material until the pressure forces due to the thermal motions act to stop it. The baryons and photons are coupled strongly in the early universe and they both contribute energy density for the gravitational attraction and pressure to counter act it. Any non-baryonic dark matter will contribute to the gravitational attraction also. The coupled system is very much like a mass on a spring. It behaves as an oscillator and there are acoustic oscillations for perturbations on all physical scales.

The oscillations are characterized by the sound speed which is essentially $1/\sqrt{3}$ the speed of light until decoupling of the photons and baryons at the surface of last scattering. This means that perturbations larger than the scale of the horizon at last scattering are still in the first compression stage with a compression roughly proportional to the physical scale over the horizon scale.

Perturbations that are nearly the horizon scale at the surface of last scattering are just maximally compressed. They are slightly smaller than the horizon size because the speed of sound will average to slightly less than the speed of light over $\sqrt{3}$. Smaller perturbations will vary from being maximally compressed to maximally rare and back to maximally compressed depending upon their physical size which determines how many oscillations have taken place. As we know from oscillator theory dissipation will cause the amplitude to die down proportional to the number of oscillations. There is dissipation due to photon viscosity and leakage.

These acoustic oscillations can be viewed as sound waves, not traveling through space but in time. The oscillations all start at the same instant but oscillate with different frequencies depending their physical scale. This results in two effects: bulk motion of the baryons and photons and thus varying velocity on the surface of last scattering, and also a varying density of photons at the surface of last scattering. These two effects are 90° out of phase with each other. Note that in an oscillation the extremes of motion (turning points) are when the velocity is zero and the velocity is highest in the middle of the oscillation.

### Receiver Motion Doppler Effect

We consider the effect of emitter and receiver motion. First consider receiver motion as illustrative of the Doppler effect and then we will consider the velocity variations at the surface of last scattering. Receiver motion with velocity $\beta = v/c$ relative to an isotropic Planckian radiation field of temperature $T_o$ produces a Doppler-shifted temperature

$$\begin{aligned} T(\theta) &= T_o \frac{(1-\beta^2)^{1/2}}{(1-\beta\cos(\theta))} \quad (34) \\ &\approx T_o\,(1 + \beta\cos(\theta) + (\beta^2/2)\cos(2\theta) + ...) \quad (35) \end{aligned}$$

The first term is the monopole CBR temperature without a Doppler shift. The term proportional to $\beta$ is a dipole, varying as the cosine of the angle between the velocity and the direction of observation. The term proportional to $\beta^2$ is a quadrupole, varying with cosine of twice the angle with amplitude reduced by $1/2\,\beta$ from the dipole amplitude.

The dipole anisotropy was detected a little more than a decade after the CMB was discovered (Smoot et al. 1977). The COBE DMR maps clearly show a dipole distribution consistent with a Doppler-shifted thermal spectrum (Kogut et al. 1993, Fixsen et al. 1993). The DMR finds the same thermodynamic amplitude for all three frequencies. Using the DMR direction and FIRAS spectrum one finds the difference in spectra taken near the hot and cold poles is described to better than 1% of its peak value by the difference in blackbody spectra. The DMR maps are well-fitted by a dipole cosine dependence. There is annual modulation by the 30 km/s earth orbital velocity. All observations are consistent with the kinematic effect of our motion relative to the CMB rest frame and thus a Doppler shift origin.

The implied velocity for the solar-system barycenter is $\beta = 0.00123 \pm 0.00003$ where we assume a value $T_0 = 2.726$ K, towards $(\alpha, \delta) = (11.17^h \pm 0.03^h, -6.7° \pm 0.3°)$, or $(l, b) = (264.4° \pm 0.3°, 48.4° \pm 0.5°)$. This in turn implies a velocity for the Galaxy and Local Group of galaxies relative to the CMB. The derived velocity is $v_{LG} = 627 \pm 22$ km s$^{-1}$ toward $(l^{II}, b^{II}) = (276° \pm 3°, 30° \pm 3°)$.

For a power-law density power spectrum $P(k) \propto k^n$ (the potential power spectrum goes as $\Phi_k^2 \propto k^{n-1}$ so that in this notation $n = 1$ is scale invariant), the contribution to the rms peculiar velocity $\langle v^2 \rangle^{1/2}$ arising from primordial potential fluctuations of length scale larger than $\lambda$ is

$$< v^2 >^{1/2} \approx 400 \text{km s}^{-1} \frac{Q}{16\mu K} \left(\frac{50 h^{-1} \text{Mpc}}{\lambda}\right)^{(n+1)/2} \quad (36)$$

where $h^{-1}$ is the $H_o$ in units of 100 km s$^{-1}$ Mpc$^{-1}$ and $Q$ is the expected average quadrupole CMB intensity amplitude. Though the primordial perturbation spectrum likely extends down to scales much smaller than 1 Mpc, dissipative effects and virialization dominate at scales of 10 Mpc and may extend out to scales of 30 to 40 Mpc. These effects on average will increase the expected rms velocity. Since the rms velocity is broadly distributed, the observed Galactic velocity of $627 \pm 22$ km s$^{-1}$ is consistent with being produced via a primordial power spectrum of perturbations at the level detected by the COBE DMR. A specific prediction of this origin is that the rms velocity flow will decrease with volume averaging size $\lambda$ according to the formula above.

The Doppler effect of this velocity and the velocity of the Earth around the Sun, as well as any velocity of the receiver relative to the Earth, is normally subtracted from the data. It is likely that some intrinsic dipole power is removed as well. Thus power spectrum investigations traditionally skip the dipole ($\ell = 1$) component.

**Last Scattering Doppler Effect**

Motion of the emitting or last scattering surface will cause CMB anisotropy via the Doppler effect. For primordial density perturbations the major expected motion is due to thermal-acoustic oscillations of the primordial baryon-photon fluid. We expect that all perturbations will undergo oscillation and that as a function of wavenumber $k$ there will be the first full compression at the sound horizon (the distance sound can propagate from the beginning of the universe until the time of last scattering). There will be alternating rarefactions and compression peaks at harmonics of the sound horizon.

We know that the velocity profile will appear as shown in Figure 4. The velocity and density curves are 90 degrees out of phase. The density enhancement wins out over the velocity in amplitude but the velocity effect causes the long slow rise to the first Doppler peak.

**Damping of Small Angular Scale Anisotropies**

Figure 5 shows the effect of damping of the oscillations including the finite thickness of the last scattering surface for a $\Omega_0 = 1$, $\Omega_b = 0.06$ and $H_0 = 50$ km/s/Mpc universe. This severe damping arises as the photons diffuse through the baryons. The diffusion is most significant at recombination, which is when the photon-baryon fluid decouples. The effect of photon diffusion is to exponentially, $e^{-\tau}$, damp the temperature fluctuations coming from the density fluctuations. There are two parts to the process: (1) diffusion of photons out of the overdense regions which reduces the temperature contrast, and (2) rescattering of the diffusing photons which tends to isotropize them. The diffusion length grows rapidly during recombination but is in effect for a shorter time. This means that the smallest physical scales are most damped. The diffusion of photons out and into the oscillations means that the oscillation is also damped (called Silk damping) which also mostly affects the smaller perturbations up until decoupling. As the photons diffuse they still are affected by the gravitational redshift and they are rescattered and isotropized, further reducing the signal.

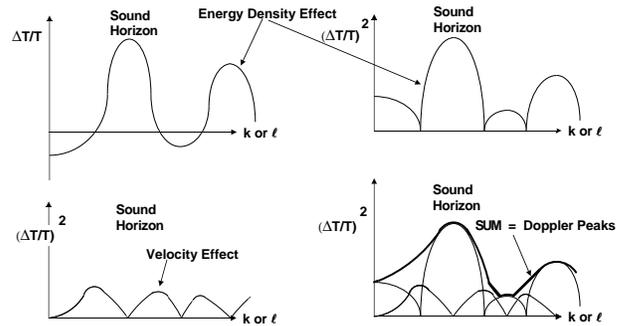

Figure 4: A rough schematic showing the contributions to the "Doppler peaks" from the density enhancement and velocities associated with the acoustic oscillations occurring up until decoupling. The upper left frame shows the temperature variation due to the change in the density of photons as a function of scale. The right frame shows the corresponding power spectrum which is just the rectified mean square power. The lower left frame shows the effect of the motion of the photon fluid during the oscillations. Note the velocity and compression/rarefaction are 90° out of phase with each other. The final panel shows the power spectra for the two components and their sum. The successive peaks and minima in the energy density are at harmonics of the sound horizon.

**The effect of re-ionization**

For a reionized universe the last scattering is delayed or stretched out, and the diffusion length grows to be nearly the horizon size at last scattering. If the universe never went through a neutral (decoupled) phase, clearly the oscillations will continue and the Doppler peaks will be shifted to the sound horizon at that last scattering time. Thus observation of the Doppler peak in the expected place can tell us that the universe did not undergo significant reionization until very late ($z < 50$) when the density of available electrons was very low so that the fluctuations survived being erased by Compton scattering. This is complementary to possible spectral measurements for checking the ionization history of the universe.

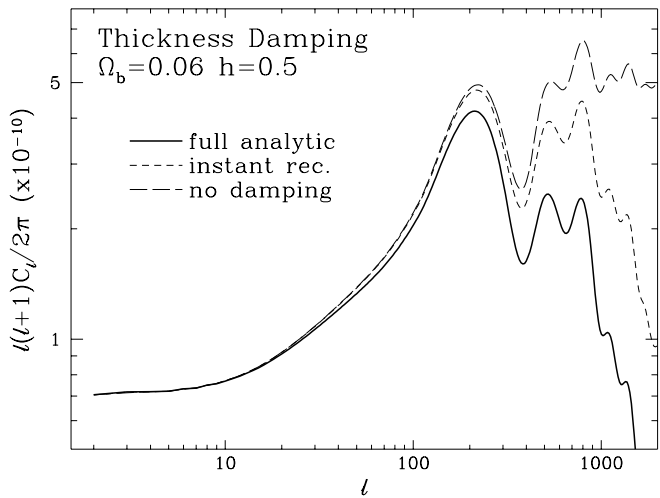

Figure 5: The effect of damping on the "Doppler Peaks". Estimating damping in the instantaneous recombination approximation leads to a significant underestimate of the damping scale (Hu & Sugiyama 1995)

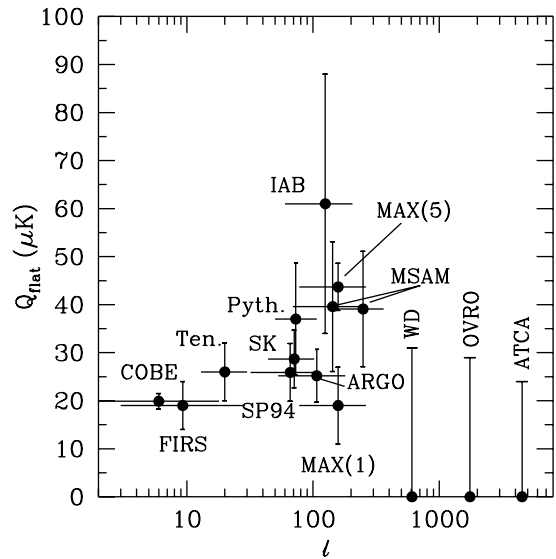

Figure 6: Current status of CMB anisotropy power spectrum observations adapted from Scott, Silk, & White (1995). The amplitudes plotted are the quadrupole amplitudes for a flat (unprocessed scale-invariant spectrum of primordial perturbations, i.e. a horizontal line) anisotropy spectrum that would give the observed results for the experiment. Figure 7 gives an indication of the expected spectrum for a processed spectrum in a CDM model for comparison.

## Anisotropy Results

Since the DMR announcement of the discovery of anisotropy nine groups have reported CMB anisotropies. Figure 6 shows the current observational status of the CMB anisotropy power spectrum.

The 'MIT' FIRS experiment (Meyer et al. 1991, Page et al. 1990) is the only experiment, other than the DMR, to map a significant portion of the sky. The FIRS experiment has a $\sim 3°$ beam width and covered nearly a quarter of the sky with a single balloon flight. The FIRS data correlate well with the DMR data (Ganga et al. 1993) and show a similar power spectrum (Ganga et al. 1994) consistent with scale invariance.

The Tenerife (Watson et al. 1991) is also a large angular scale experiment (beam width 5°) that covers a differenced (8°) strip scanned on the sky by the earth's rotation. The Tenerife experiment has pointed out bumps on the sky as specific locations of anisotropy (Hancock et al. 1994). The ULISSE experiment (de Bernardis et al 1992) reported upper limits on 6° CMB anisotropy using balloon-borne bolometric observations. The Advanced Cosmic Microwave Explorer (ACME South Pole) (Gaier et al. 1992 & Schuster et al. 1993) reported upper limits and detections of fluctuations operating with HEMT amplifiers. The Saskatoon "SK93" experiment (Wollack et al. 1993) used HEMT amplifiers to detect CMB anisotropy from Saskatoon, SK, Canada. Fluctuations were reported from South Pole observations by the Python experiment (Dragovan et al. 1993). The ARGO balloon-borne experiment (de Bernardis et al. 1994) observed a statistically significant signal with a 52' beam. The Italian Antarctic Base (IAB) experiment (Piccirillo and Calisse 1993) used bolometric techniques with a 50' Gaussian beam and reports anisotropy. The Millimeter-wave Anisotropy eXperiment (MAX) is a balloon-borne bolometric instrument with high sensitivity in the medium angular scale that has completed five flights detecting significant CMB fluctuations (Alsop et al. 1992, Meinhold et al 1993, Devlin et al 1994, Clapp et al 1994). The Medium Scale Anisotropy Measurement (MSAM) balloon-borne experiment (Cheng et al. 1994) is a very similar balloon-borne medium-scale CMB anisotropy instrument but with a different chopping scheme that allows the results to be reported either as a difference or a triple difference, providing two effective window functions. Also from the South Pole the White Dish experiment (Tucker et al. 1993) reports an upper limit on CMB anisotropy. Arc-minute scale anisotropy upper limits were reported using the Owens Valley Radio Observatory (OVRO) (Myers et al. 1993). The Australia Telescope Compact Array (ATCA) was used to place upper limits on CMB anisotropy in a Fourier synthesized image (Subrahmayan et al. 1993).

The field is moving so rapidly that such plots get out of date quickly. At this stage we can begin to see that the reported results, though scattered, are actually in rough agreement with each other and with many models. The goal at the moment is to refine the results and our ability to distinguish between models. It is a hot topic whether the data show evidence for a Doppler peak and then whether that peak is in the right location. The existence of the peak and its location would go far towards telling us not only whether we are on the right track with these models, and if there is a connection with large scale structure formation, but also about various cosmological parameters.

## §3 Instruments

We can anticipate significant advances over the next few years as many groups continue ground-based measurements both with the conventional beam switching and interferometer techniques, balloon-borne experiments including long-duration flight instruments BOOMERANG and TOPHAT as well as the idea of ultra-long duration balloon flights, e.g. ACE. There is also significant effort going into the design of potential space missions including COBRAS/SAMBA, PSI, MAP, and FIRE. In looking toward the future, it is instructive to consider these instruments and their likely results. Their likely progress drives both the field and the design of and need for future work. (The space missions are considered in the FUTURE section.)

### 3.1 MAX/MAXIMA

Consider the MAX/MAXIMA payload as representative of current and currently planned balloon-borne missions.

**MAX**

The Millimeter-wave Anisotropy eXperiment (MAX) is a balloon-borne bolometric instrument which observes at multiple frequencies with high sensitivity on the 0.5° angular scale. MAX has completed five flights detecting significant CMB fluctuations (Fischer et al. 1992, Alsop et al. 1992, Meinhold et al 1993, Devlin et al 1994, Clapp et al 1994, Tanaka et al. 1995, Lim et al. 1995).

The MAX instrument consists of an off-axis Gregorian telescope and a bolometric photometer mounted on an attitude-controlled balloon-borne platform which makes measurements at an altitude of 36 km. The Gregorian telescope consists of a 1-meter primary and a nutating elliptical secondary. The underfilled optics provides a 0.55° FWHM beam when focused and aligned. The 5.7 Hz nutation of the secondary modulates the beam on the sky sinusoidally though ±0.68° and the attitude control sweeps the beam over a 6° or 8° path and back in about 108 seconds, producing about 15 to 20 independent temperature differences on the sky. Depending upon the time of observation and location of region under observation sky rotation can cause the observed region to be in the shape of a bow-tie.

On flights 4 & 5 the single-pixel four-band bolometric receiver features negligible sensitivity to radio frequency interference and an adiabatic demagnetization refrigerator to cool the photometer to 85 mK. The dichroic photometer used for MAX has ($\delta\nu/\nu$) of 0.57, 0.45, 0.35, and 0.25 filter bands at 3.5, 6, 9, and 15 cm$^{-1}$. MAX covers the high frequency side of the window formed by galactic dust emission rising at higher frequencies and Galactic synchroton and free-free emission increasing at lower frequencies. The 15 cm$^{-1}$ channel acts as a guard against Galactic dust and atmospheric emission. The multiple frequencies provide sufficient redundancy to provide confidence that the signal is CMB and not a foreground or systematic effect.

MAX is calibrated both by an on-board commandable membrane and by observations of planets, usually Jupiter. The two techniques agree at roughly the 10% level. The calibration is such that the quoted temperature difference is the real temperature difference on the sky.

MAX makes deep CMB observations (typically one hour) on regions generally selected to be low in dust contrast and total emission and free from known radio sources. MAX has made observations on five flights The data from most of the scans are in good agreement but the scan of the mu Pegasi region is significantly lower than the rest. That is why there are points plotted for the average of the agreeing region and a single point for the mu Pegasi data. We are hopeful that the 5th flight will resolve this issue but it seems to be coming out at an intermediate value.

The center of the scan is the same for the three observations of GUM (the star Gamma Ursae Minoris) but the relative geometry is such that the three scans made bow-tie patterns which cross at the star. White and Bunn (1995) have made use of this fact to construct a two dimensional map of the region which is roughly 10° × 5°. The title of their paper is "A First Map of the CMB at 0.5° Resolution".

Making maps is clearly the appropriate approach for the current generation of new experiments. MAX is evolving to a new system MAXIMA, which is designed and constructed for the goal of getting the power spectrum around the first "Doppler" peak and making maps covering a significant portion of the sky.

**MAXIMA**

MAXIMA stands for MAX imaging system. The current one-dimensional scans are very useful data for the dis-covery phase of CMB anisotropy research. Soon progress will depend upon the availability of two-dimensional maps of low galactic foreground regions (low dust in this case) with several hundred pixels so that sampling variance is less important (see section 4.1). In addition one can look for properties of the sky which are not predicted by theories and could be overlooked in statistical analyses. It also makes it possible to catalog features for comparison to or motivation of other experiments.

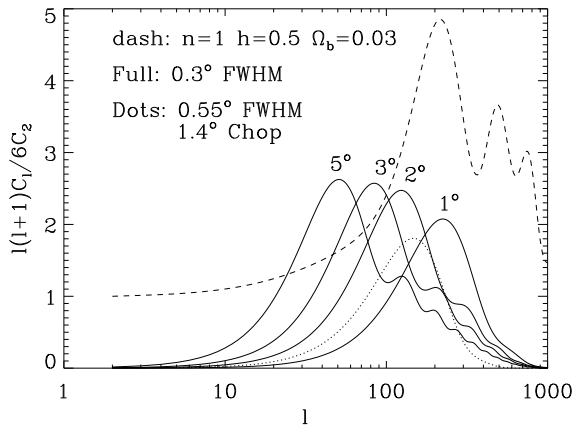

Figure 7: Window functions possible for the new MAXIMA mirror for chop angles of 1, 2, 3, and 5 degrees and the dotted line shows the MAX window function. For reference a theoretical power spectrum for CDM is shown as a dashed line. Figure courtesy Shaul Hanany.

Under the auspices of the NSF Center for Particle Astrophysics a collaboration consisting of groups from the University of California at Berkeley, Caltech, the University of Rome, and the IROE-CNR Florence have begun work on a new system. To make an imager a new optical system was necessary. The primary feature is a 1.3-meter, off-axis, light-weight primary mirror. The primary will be modulated which allows a much larger beam chop angle on the sky with less spill over and thus more pixels in the focal plane. Cold secondary and tertiary mirrors provide a cold Lyot stop and the field-of-view required for the array of 20 arcminute pixels. The geometrical aberrations in the center of the field-of-view are less than 10 arcminutes.

A larger primary mirror requires a larger gondola which is now constructed. The chop angle can both be increased and varied allowing the instrument to sample the shape of the power spectrum over the range $40 < \ell < 300$. Figure 7 shows some sample window functions possible with the new system.

An additional feature is new detector electronics with AC coupling in order to allow linear scanning in a total power mode, making maps and power spectrum measurements directly. This approach is different than that of making a number of different window functions as shown in figure 7. The idea is to use a scan or raster scan of the CMB anisotropies on the sky directly rather than obtaining a set of differences at different chop angles. One is thus mapping directly and measuring the power spectrum as the fourier transform of the data. At this stage the instrument is designed to operate in this mode either by scanning the primary mirror in a sawtooth pattern or moving the entire gondola in azimuth.

Another major change will be going from a single pixel four-frequency photometer to an eight-pixel receiver. This will allow taking data at eight times the rate and thus make two-dimensional mapping feasible. The receiver design has been completed and the new dewar ordered. The bolometers will be changed to have a spide-web substrate so that cosmic ray transient occurences will be reduced by more than an order of magnitude. The target date of a first flight of the new gondola is August 1995. We can anticipate that within three years MAXIMA will have made maps and will have measured the anisotropy power spectrum around the location of the first doppler peak.

## 3.2 MSAM/TOPHAT

MSAM/TOPHAT is similar to MAX/MAXIMA at the present. A notable difference between MSAM and MAX has been that MSAM used a three-position chop analyzed either as a triple beam or double beam (two chop angles on the sky) observation (Page et al. 1994). MSAM angular resolution is $0.5°$ between 5 and 23 $cm^{-1}$ (150 and 700 GHz or wavelengths 0.4 to 2.0 mm). MSAM has had two flights (June 1992 and May 1994) both from Palestine, Texas.

MSAM is preparing for another flight with expanded frequency coverage in 5 spectral bands between 2.3 and 5 $cm^{-1}$ (70 to 150 GHz or wavelengths 2.0 to 4.3 mm). The instrument is expected to improve its signal-to-noise ratio by about a factor of three over the previous results.

All observations are along a ring surrounding the north celestial pole.

TOPHAT is conceived as a long-duration balloon-borne experiment with the detectors located on the top of the balloon rather than in a gondola hanging below the balloon. It will observe in five spectral bands between 5 and 21 $cm^{-1}$ (150 and 630 GHz or wavelengths between 0.5 to 2.0 mm). The current plans call for the measurement of 40 points on the sky, each with an rms sensitivity of $\delta T_{rms} \approx 1\ \mu K$ or $\delta T_{rms}/T_{CMB} \approx 3 \times 10^{-7}$ including removal of the galactic foreground dust emission. I predict that TOPHAT will evolve towards a less sensitive observation of a larger area of the sky for reasons that will be discussed in the limitations section. It will be useful to have some deep (high-sensitivity) scans such as this or what MAXIMA can do to understand the signal well. In this case sampling variance and other science drives one toward observing a larger fraction of the sky.

## 3.3 BOOMERANG

BOOMERANG is to first order the long-duration balloon-borne version of MAXIMA and intermediate step toward a bolometer space mission, e.g. FIRE or COBRAS/SAMBA. It is planned to make a many day flight circumnavigating Antarctica as early as December 1996. BOOMERANG will move more directly towards mapping a significant region of the sky than the experiments thus far discussed.

## 3.4 ACE

As a follow up to their South Pole HEMT observations the Santa Barbara group has proposed ACE (Advance Cosmic Explorer). It is a large, light-weight (200 kg), system aimed at making flights lasting 90 days or more. It would utilize advanced HEMTs, active refrigerators, and a 2-m diameter composite mirror to cover the frequency range 25 to 90 GHz. In three such flights such a system could map 75% of the sky to an angular resolution of 10 arcminutes at a level of about 20 $\mu$K. This project is still in the early phase but is indicative of what with sufficient funding one might achieve by the year 2000.

## 3.5 Ground-Based Instruments

Ground-based instruments have made a significant contribution to CMB anisotropy observations. They have been more successful than originally envisioned as a result of the observers' clever strategies to minimize and reduce the effect of the atmosphere. These strategies have included going to high, dry sites such as the South Pole and Teide peak on Tenerife and using triple-beam chopping or other similar techniques. These techniques are more difficult to use when going to mapping and making observations over an extended portion of the power spectrum. Here again it is possible that significant progress can be made though it is likely to be eventually limited before the science is exhausted.

An exciting exception is the use of aperture synthesis interferometers. The Ryle Telescope images of the Sunyaev-Zeldovich effect in clusters and the CAT (Cambridge Anisotropy Telescope) results have convinced many that interferometers have a bright future in actually mapping anisotropy on small angular scales over selected regions of the sky. A number of proposals are pending. Noteworthy are the VSA (Very Small Array) in England and the Caltech interferometer. If funded, these interferometers are likely to provide a very good first cut at the CMB anisotropy power spectrum on angular scales less than about 0.5° ($\ell$>400).

## §4 Limitations: Cosmic Variance, Sensitivity, & Foregrounds

The fundamental limitations to these observations are the cosmic variance, sensitivity of the detectors and the unavoidable foregrounds. Systematic errors are a difficult and thorny issue and must be dealt with carefully. That it is possible to overcome systematics has an existence proof in the form of accepted results to date. The difficulties are not to be minimized and improving by an order of magnitude is likely to uncover new issues. However, we shall assume that observers will be able to overcome these as they are not necessarily fundamental limitations.

## 4.1 Cosmic & Sample Variance

An important effect that must be considered in understanding the temperature fluctuations expected and how accurately one can determine fundamental parameters is cosmic and sampling variance. In most cosmological models, the observed CMB temperature field is a single realization of a stochastic process. A single realization, e.g. the observable universe set by our horizon, will not, in general, exactly follow the ensemble mean of the parent population ("cosmic variance"). If we only cover a portion of the sky, we do less well and we are limited by "sample variance". This problem is most acute at the largest angular scales; the CMB quadrupole, for example, is described by only 5 parameters. At higher multipole moments the larger number of components ensures that cosmic variance becomes less important. Incomplete sky coverage can introduce another uncertainty, that the fraction examined may not be representative of the realization as a whole ("sample variance"). Sample variance is not an issue for the full-sky *COBE* maps, but it is an important limitation for observations at smaller angular scales.

In general the cosmic variance adds an irreducible theoretical uncertainty to each spherical harmonic amplitude with a variance equal to twice the variance of the spherical harmonic amplitude. For the power spectrum this means that the cosmic variance rms error at each $\ell$ is equal to the power at $\ell$ times the factor $\sqrt{2/(2\ell+1)}$. This is quite a serious uncertainty for low $\ell$. Figure 8 shows the limitation set by cosmic variance (i.e. assuming full sky coverage with uniform accuracy) in determining a sample power spectrum.

If the sky is not covered uniformly or if only a portion of the sky is sampled, then the sample variance must be larger than the cosmic variance. In general the increase in variance is set by the ratio of the whole sky to experiment effective solid angles (Scott, Srednicki, & White 1994). It is appropriate to note that at present sample variance is as large or larger than the instrument noise and calibration uncertainty for many current balloon results.

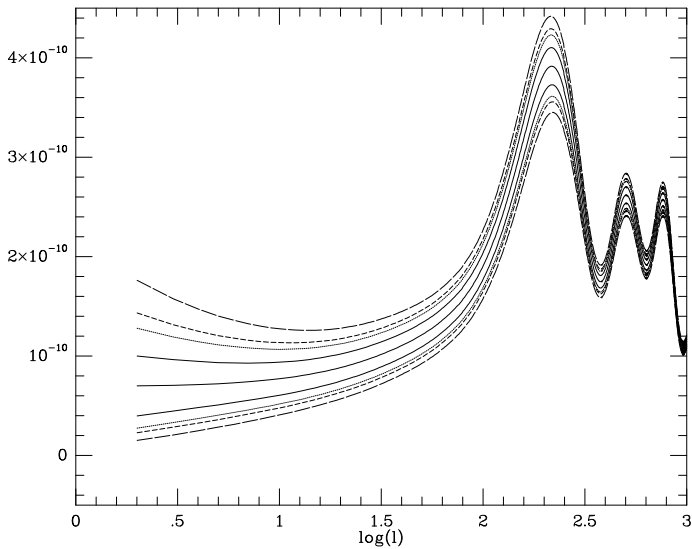

Figure 8: Center line shows the power spectrum for standard CDM ($h = 0.5, \Omega_b = 0.05$, courtsey of Sugiyama) The pairs of lines are the cosmic variance at 68%, 90%, 95%, and 99%, respectively.

## 4.2 Sensitivity

Sensitivity has always been an issue in CBR anisotropy observations. For coherent receivers (those observing a single coherent state) the rms fluctuations in output power is given by the formula

$$\delta T_{rms} = \frac{s(T_{system} + T_{obs})}{\sqrt{B \tau}} \quad (37)$$

where $T_{system}$ is the equivalent noise power of the system expressed in units of antenna temperature $P = kT_A B$, $T_{obs}$ is the antenna temperature of the observation target – which when looking at the CBR is often about 5 K which is set by the CBR 3 K and the foreground input noise power. $B$ is the bandwidth, $\tau$ is the observation time, and $s$ is the sky observation scheme and is a factor near unity (typically $1/\sqrt{2} \leq s \leq 2$ to $\pi$).

Technology has been the usual limit for coherent receivers but recently the system temperatures have been getting noticeably smaller $10 \leq T_{system} \leq 100$ K. The future promises to provide receivers that will have $T_{system} \leq 10$ K which leaves increasing the bandwidth as the only means to improve sensitivity of a given receiver. That too is limited for practical and observational reasons. The only avenue left is to go to multireceiver systems - e.g. arrays. A good system with 10 K system temperature, a 5 K sky temperature and a 10% bandwidth would have the rms sensitivity and integration time to $\delta T/T = 10^{-6}$ (including the conversion from differential antenna temperature to thermodynamic temperature) shown in the following table:

### Sensitivity & Observation Times

| Frequency (GHz) | Wavelength (cm) | Sensitivity ($\mu$K $s^{1/2}/t^{1/2}$) | Time to $10^{-6}$ (sec) |
|---|---|---|---|
| 30 | 1 | 274 | 10500 |
| 45 | 0.667 | 237 | 7500 |
| 53 | 0.567 | 206 | 6600 |
| 90 | 0.333 | 158 | 5100 |
| 125 | 0.24 | 134 | 5300 |
| 150 | 0.20 | 122 | 6100 |
| 210 | 0.143 | 104 | 11400 |
| 300 | 0.1 | 87 | 50000 |

These times are very long in that they require on the order of an hour or more per pixel and one needs many pixels for reasons of statistics. This means that to achive this sensitivity level over a significant portion of the sky one must utilize both arrays of detectors and one of three approaches: (1) ground-based observations (probably interferometers will prove to be the best approach because of the atmospheric foreground problem), (2) very long duration ballooning, or (3) space-borne (satellite dedicated to these observation) platforms.

Incoherent detectors are the other approach to observing the CBR. They generally work by measuring the total power absorbed (bolometers) and thus can look at multiple modes (quantum states). They gain sensitivity by utilizing a very wide bandwidth, multiple modes, and by cooling the detectors to very low temperatures, typically $\leq 0.3$ K. Filter technology and optical efficiency for such systems is not yet a fully mature technology though moderately so. Thus the bandpass and optical efficiency are not likely to be well-described by a square or simple function. The sensitivity of a bolometer is difficult to quantify as precisely as for a coherent receiver since it depends upon the particular configuration, especially the bandpass filter and optical efficiency. The sensitivity of a good bolometer system these days is roughly $200\mu$K $s^{1/2}$ thermodynamic temperature near the peak of the spectrum and falling to either side as the power decreases due to either the high-frequency Wien fall off and the decrease in states and energy at lower fequencies. Ultimately it may be possible to make bolometer detector systems with up to a factor of 10 improvement over this sensitivity.

At the present time, the best detectors (both coherent and bolometers) are roughly comparable in sensitivity but separated in frequency. Generally coherent recievers work better at lower frequencies $\leq 100$ GHz. and bolometers work better at higher frequencies $\geq 80$ GHz.

It is interesting to compute the fundamental limit to anisotropy observation sensitivity. It is easiest to understand in terms of photon counting as there are the least number of complicating factors to consider. However, one cannot just take the 415 photons per cubic centimeter arriving at the speed of light multiplied by collecting area ($5.17 \times 10^{15} \times$ Area) to find the number of photons counted. A quantum-limited detector in the sense of one that counts the photons incident and adds no other noise is the best one can do as a detector. However, one must also collect those photons onto the detector. For a coherent detector it is obvious that the number of photons detected per second is set by the integral over the bandwidth, $B$, of the occupation number and that one collects $B$ such states per second.

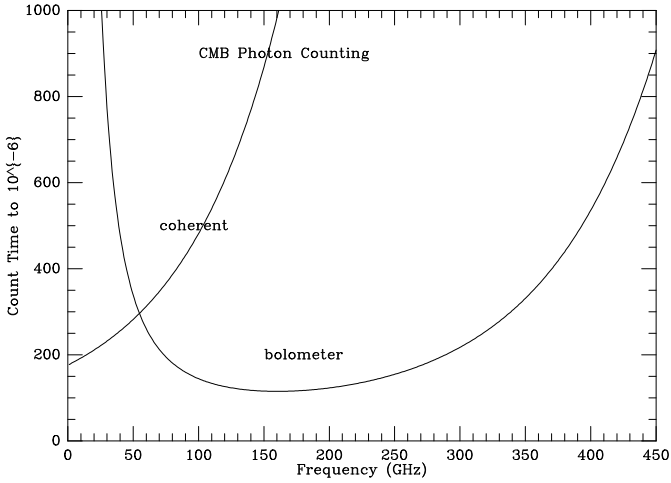

Figure 9: The time to a sensitivity of $\delta T_{rms}/T_{CMB} = 10^{-6}$ set by detectors that count the individual CMB photons. The upper line is for an ideal coherent detector with a 10% bandwidth. The lower line is for an ideal photon counter (e.g. bolometer) with throughput $A\Omega = 0.3$ cm$^2$ steradian and ideal optical efficiency over a 10% bandwidth. The time for bolometers is likely to rise more rapidly as frequency increases as some systems will obtain higher angular resolution at higher frequency at the expense of $A\Omega$.

$$N_{coherent} = B \int \epsilon(\nu) n(\nu) d\nu = B \int \frac{\epsilon(\nu)}{e^{h\nu/kT_o} - 1} d\nu \quad (38)$$

where $\epsilon(\nu)$ is the optical efficiency of the system at each frequency and $B = \int \epsilon(\nu) d\nu$. The limit is easily understood in terms of thermodynamic equilibrium. The power available at the terminals of a resistor is simply $P = kT_aB$. If a such a resistor is connected to an antenna system, and if it is looking at a blackbody of temperature $T_o$, it must come to the same temperature and emit and receive power at the same rate. The number of photons received per unit time is then just $N = kT_aB/h\nu = B/[exp(\frac{h\nu}{kT_o}) - 1]$ which is the same as before. There is a conspiracy set by diffraction that means the solid angle times collecting area is just the wavelength squared $\Omega A = \lambda^2$.

Figure 9 shows the minimum time to a sensitivity of $10^{-6}$ for an ideal coherent detector that counts photons and for an ideal bolometer.

For bolometer systems the argument is slightly more involved. It is the collection of photons onto the detector that limits the number of photons actually available to be counted. However, now we can not use the diffraction relationship $\Omega A = \lambda^2$. We must actually determine the throughput (étendue) and optical efficiency. The throughput is set by a number of considerations such as the size of the optics but it can be made larger than $\lambda^2$ by a substantial amount. The number of photons per unit bandwidth per unit time is then the brightness times $A\Omega$ times the optical efficiency:

$$N = \frac{2}{exp(\frac{h\nu}{kT_o}) - 1} \frac{A\Omega}{\lambda^2} \epsilon(\nu) = 2N_{coherent} \frac{A\Omega}{\lambda^2} \epsilon(\nu) \quad (39)$$

The factor of two comes from the fact that two polarizations are available. Coherent detectors can gain back the polarization factor of 2 by putting two detectors on the same photon collecting optics. The advantage of bolometers (incoherent detectors) is the polarization factor of two and the possibility of making the throughput $A\Omega$ significantly larger than $\lambda^2$.

Bolometers retain the advantage of throughput. Generally for a simple optics system the throughput is constant for frequencies above the lowest modes, since geometrical optics is more relevant than diffraction. The lowest (optics cut off) mode and area is set by the collecting size of the bolometer. For example, on the 5th flight of MAX the throughput was roughly 0.12 cm$^2$ steradian.

At present the size (collecting area) of the bolometers is limited by the heat capacity and conductivity and by cosmic ray hits. Improvements such as the "spider" absorber used by MAXIMA and BOOMERANG may allow bolometers to overcome the current limits significantly. If so, then bolometers may be able to push their performance margin over coherent receivers such as HEMTs down to frequencies as low as 40 GHz.

One can estimate the combined effect of cosmic (or sample) variance and instrument noise relatively simply given that one expects them to be random and independent of each other. To first order the random fluctuations can be treated as gaussian and the variance is (see e.g. Knox 1995)

$$\left\langle (C_\ell^{obs} - C_\ell)(C_{\ell'}^{obs} - C_{\ell'}) \right\rangle = \frac{2}{2\ell + 1} \left( C_\ell + <\sigma_{lm}\sigma_{l'm}>_m \right)^2 \quad (40)$$

where $\sigma_{lm}$ is the error in the spherical harmonic coefficient $a_{lm}$. Conceptually this is just

$$\frac{\Delta C_\ell}{C_\ell} = \sqrt{\frac{2}{2\ell + 1}} \left( 1 + \frac{1}{S/N_\ell} \right) \quad (41)$$

where $S/N_\ell$ is the mean signal-to-noise for an $a_{\ell m}^2$, i.e. $S/N_\ell = \sigma_{\ell m}^2 / a_{\ell m}^2$, since $C_\ell \equiv \left\langle a_{\ell m}^2 \right\rangle_m$.

If the process that causes anisotropies is stochastic with random phase, there is a fundamental limitation set by the cosmic or sample variance. One gains little in determining the power spectrum by making observations with $S/N_\ell$ better than about 10. For a sensitivity limited experiment, generally the signal to noise $S/N_\ell$ improves with the inverse of the observation time.

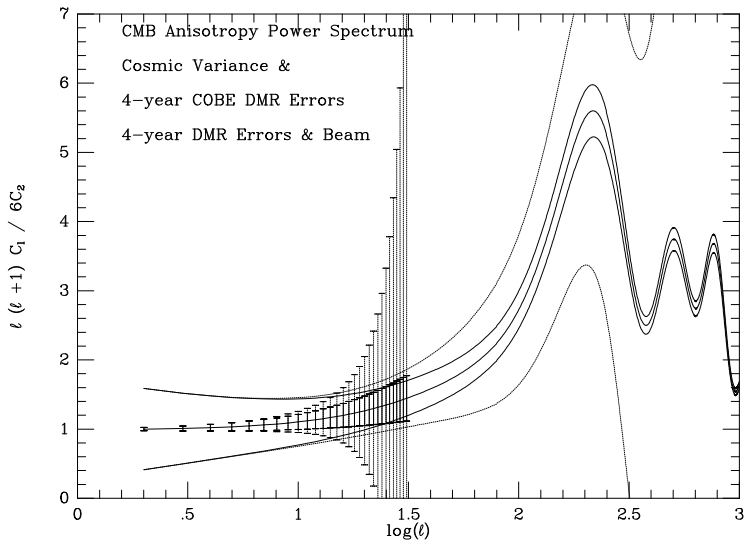

Figure 10: Center line shows the power spectrum for standard CDM ($h = 0.5, \Omega_b = 0.05$, courtsey of Sugiyama) The pair of lines on either side show the effect of cosmic variance alone. The solid error bars show the effect of instrument noise for the COBE DMR 4-year 53 GHz data set alone. The dotted error bars show the effect of DMR instrument noise combined with the beam pattern. The dotted pair of lines shows the anticipated combined effect of cosmic variance and instrument noise without the effect of the beam filter. This shows the importance and the gain from having as high a resolution as possible. It also shows that future experiments will require better net sensitivity than the DMR by more than a factor of 10.

If less than the full sky is measured then there will be a larger error because fewer samples are taken, to first order one has

$$\frac{\Delta C_\ell}{C_\ell} = \sqrt{\frac{2}{2\ell+1}} \sqrt{\frac{4\pi}{\Omega_{obs}}} \left(1 + \frac{1}{S/N_\ell}\right) \qquad (42)$$

where $\Omega$ is the effective solid angle of the sky covered. (Effective means weighted by the total error: sampling and noise.)

Or equivalently, if each pixel has the same error $\sigma_{pix}$

$$\frac{\Delta C_\ell}{C_\ell} = \sqrt{\frac{2}{2\ell+1}} \sqrt{\frac{4\pi}{\Omega_{obs}}} \left(1 + \frac{4\pi \sigma_{pix}^2}{W_\ell C_\ell \times N_{pix}}\right) \qquad (43)$$

where $N_{pix}$ is the total number of pixels observed and $W_\ell$ is the beam filter function. Clearly one wants $W_\ell$ as close to unity as possible for the $\ell$'s of interest. If the sky coverage is not uniform, then an effective weighting by determines the signal-to-noise term which is then approximately $\sigma_{obs}^2/(C_\ell \times t_{total} \times f)$ where $\sigma_{obs}$ is the detector noise per second, $t_{total}$ the total observation time, and $f \equiv \langle Y_{\ell m} Y_{\ell m}^* \rangle$ which represents the average of the $\ell$-spherical harmonics weighted by the sky coverage.

The end conclusion is that for a fixed observation noise and time, the measurement of the power spectrum is improved by covering as uniformly as possible as much sky as possible with as much angular resolution as possible.

There may be credibility and observational reasons for getting a higher signal-to-noise per pixel. However, great concentration on a small portion of the sky only allows one to understand the signal and systematics including possible foregrounds better. If the systematics and foregrounds can be neglected, then the optimum is clearly biased to greater sky coverage.

It turns out also to be true that higher angular resolution also improves the measurement of the power spectrum substantially. This is especially true when one considers fitting the power spectrum. Figure 10 shows this clearly with the anticipated limitation set by the instrument noise and beam size and cosmic variance in determining a sample power spectrum for the full four-year COBE DMR data set. For foreground and practical reasons the bandwidth is usually limited to roughly 10% for coherent receivers and 20 to 30% for bolometer systems. The frequency range that is likely to be most optimal for CBR anisotropy observations is the range $50 \geq \nu \geq 160$ GHz with the best depending upon angular scale of observation. This window is set by unavoidable foreground emissions.

### 4.3 Foreground Emissions

The microwave sky is dominated by relic emission from the cosmic microwave background and local emission from within our Galaxy. The spatial distribution and frequency spectrum of these components probe physical conditions and processes ranging from the early universe to the local interstellar medium.

Figure 11 shows the spectra of the CMB and Galactic emissions. Galactic emission at millimeter wavelengths is dominated by thermal emission from interstellar dust. Galactic emission at centimeter wavelengths is dominated by two components: synchrotron emission from cosmic-ray electrons accelerated in the Galactic magnetic field, and free-free emission (thermal bremsstrahlung) from the warm ($T_e \sim 8000$ K) ionized interstellar medium. Emission from Galactic sources probes physical conditions in the interstellar medium. An understanding of the combined Galactic foregrounds tests models of the large-scale structure and energy balance of the Galaxy, and is crucial to mapping the CMB spectrum and anisotropy.

Three phases of the interstellar medium are known to co-exist in the Galaxy: a cool phase ($T \lesssim 100$ K) in neutral clouds, a hot ($T \sim 10^6$ K) "coronal" component, and a warm ($T \sim 10^4$ K) ionized component. The cool component, including both dense and diffuse molecular clouds, comprises no more than 2% of the ISM by volume. The volume partition between the hot and warm ionized gas, central to an understanding of the energy balance of the Galaxy, is unknown. Possibilities consistent with existing observational data range from a quasi-homogeneous warm background with isolated cavities or superbubbles of hot gas, to a rough equipartition between the components, to a pervasive and dominant hot coronal medium. The coexistence of the phases is governed by the balance between heating and cooling mechanisms; however, the heating mechanisms of the respective media are poorly understood (Reynolds & Cox 1992). Our knowledge of the ionized ISM derives from dispersion measurements toward pulsars, radio continuum measurements, atomic recombination lines, and methodical observations of H$\alpha$ and various optical forbidden lines (Reynolds 1990). Much remains to be learned about the ionized interstellar gas – notably, the sources of its ionization far from the disk, its heating, and its distribution throughout the Galaxy. The ionization has been ascribed to ultraviolet flux from the young stellar population of the disk, shocks, Galactic fountain flows, or decaying dark matter. Observed column densities of highly ionized species (Savage & Massa 1987) support a picture involving collisional ionization from a Galactic fountain-type flow together with photoionization either by an extragalactic radiation field or from the disk, or else by ionizing radiation emitted by cooling gas in a Galactic fountain. Ionization from O-star associations within isolated cavities (superbubbles) due to supernovae explosions is also consistent with current observations.

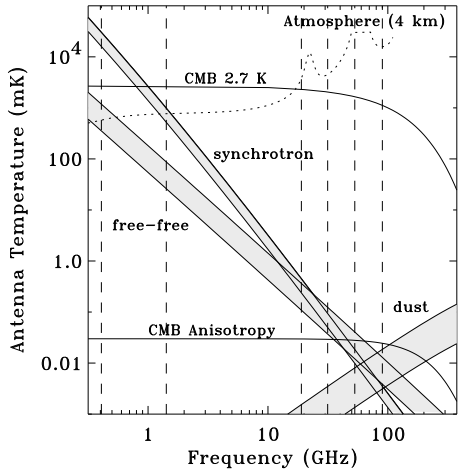

Figure 11: CMB and foreground emission spectra. The shaded regions indicate the range of synchrotron, free-free, and dust emission at Galactic latitude $15° < |b| < 75°$. The dotted line indicates typical atmospheric emission from a mountain or Antarctic site. Solid lines indicate the mean CMB spectrum and rms amplitude of anisotropy. Vertical dashed lines indicate published sky surveys, including *COBE*. There are no sensitive surveys at cm wavelengths.

### Intellar Dust Emission

Interstellar dust hides much of the Galactic plane in the optical. Longer wavelength observations are able to penetrate this obscuration, allowing study of individual objects. The obscuration at high Galactic latitudes is two orders of magnitude less but the signals sought by extragalactic astronomers and cosmologists are typically more diffuse and subtle, so high latitude dust remains a concern. In some regions of the sky dust emission can be a significant signal in the mm-wavelength range and affect cosmic microwave background (CMB) observations. As optical infrared measurements have moved deeper into the infrared towards longer wavelengths, the cooler components of interstellar dust have been revealed. In cooler regions dust has been observed at typical temperatures ranging from 200 K down to levels near 20 K. Dense, cold molecular clouds are expected to contain dust at temperatures as low as a few Kelvin.

Cosmologists have long worried that there might be a significant component of cold cosmic or Galactic dust that is affecting and contributing to the observed cosmic signals. We are now beginning to gather sensitive measurements with sufficient wavelength coverage to explore the high latitude dust. So far a consistent picture can be drawn from the observations. There is no evidence for a significant component of interstellar dust colder than 15 K at high Galactic latitudes. However, there is evidence that the emissivity of the dust is not as steep as frequency squared but more like $\nu^{1.5}$ in the mm-wavelength range. There is, in fact, evidence against high-latitude dust in the temperature range $4 < T_{dust} < 15$ K (Smoot 1995).

The good news is that the level of dust emission is sufficiently low as to allow observations of CMB anisotropy to the $10^{-6}$ level over a reasonable frequency range. Often plots are made showing the confusion versus frequency for various angular scales (e.g. Figure 13). Figure 12 shows the power spectrum of dust emission as observed by the COBE DIRBE instrument. Various regions of the sky were selected and the power spectrum computed and compared. The result is as anticipated. The interstellar dust emission has more power on large angular scales than on small to the extent that even plotted as $\ell(\ell+1) \times T_\ell^2$ (for direct comparison with the CMB power spectrum) the dust power spectrum decreases with increasing $\ell$.

### Galactic Synchrotron Emission

Synchrotron emission from relativistic electrons accelerated in the Galactic magnetic field dominates the Galactic foreground at long wavelengths. The volume emissivity for a power-law distribution of electrons, $N(E) =$

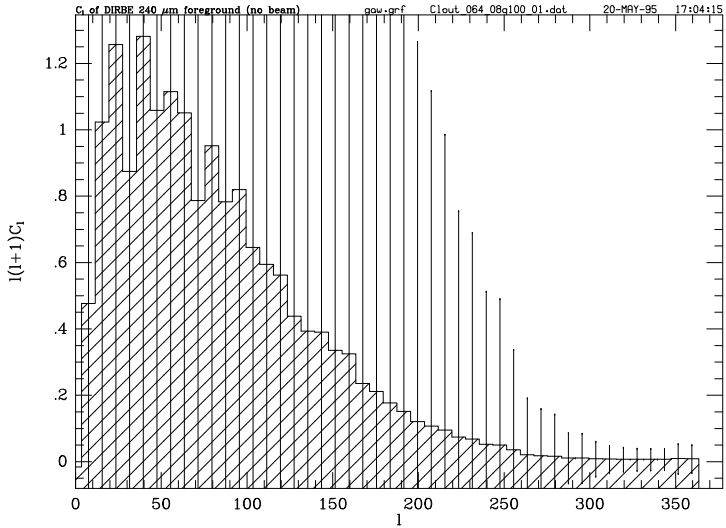

Figure 12: The interstellar dust power spectrum as observed by DIRBE at 240 microns. The errors indicate the scatter in signal from one region to another. The scatter is large because most power is on large angular scales. The signal level is sufficiently low to allow precise CMB anisotropy observations.

$AE^{-p}dE$, is

$$\epsilon(\omega) = \frac{\sqrt{3}Ae^3B}{2\pi m_e c^2(p+1)} \Gamma(\frac{p}{4}+\frac{19}{12}) \Gamma(\frac{p}{4}-\frac{1}{12}) [\frac{m_e c\omega}{3eB}]^{-\frac{p-1}{2}} \quad (44)$$

where $B$ is the magnetic field intensity. At centimeter wavelengths, synchrotron emission can be approximated as a power-law in antenna temperature

$$T_{\text{synch}} \propto \nu^{\beta_{\text{synch}}} \quad (45)$$

where the spectral index $\beta_{\text{synch}} \approx -2.8 \pm 0.3$ is a function of the electron energy spectrum and magnetic field intensity, and will vary across the sky. The synchrotron spectral index is a local approximation only: over large frequency intervals, the curvature of the synchrotron spectrum can not be ignored, while spatial variations are important even in a restricted frequency range. Bennett et al. (1992) present a model of synchrotron emission based on radio surveys at 408 and 1420 MHz and local measurements of the cosmic ray electron energy spectrum. Existing sky surveys, including *COBE*-DMR, provide only a weak test of this model. Systematic uncertainties and noise combine to limit direct knowledge of the synchrotron spectral index to $\pm 0.1$, insufficient to test the assumption that the local electron spectrum is representative of the Galaxy as a whole. Precise measurements at centimeter wavelengths, tied to a common calibration standard, could detect the predicted steepening of the synchrotron spectrum and map the spatial variations in $\beta_{\text{synch}}$ to precision $\pm 0.02$. Such measurements would test models of cosmic-ray acceleration through supernovae shocks and OB associations (Banday & Wolfendale 1990), and verify that the locally-measured electron energy spectrum is representative of the Galaxy as a whole.

Fortunately, the synchrotron emission falls off so steeply with frequency that it allows precise measurements at frequencies of 50 GHz and above. Again one can compute the power spectrum of the synchrotron maps and find that much of the power is on the largest angular scales. The frequency dependence is sufficiently different from the CMB's that separation is relatively easy.

**Free-Free Emission**

As a Galactic foreground free-free emission arises from the thermal collisions in the interstellar plasma. For temperatures below $T_e < 10^6$ K and frequencies below $\nu < 10^{10} T_e$ Hz, which covers the entire CMB spectrum for most interstellar plasma, the optical depth is

$$\tau \approx 0.08235 T_e^{-1.35} \nu^{-2.1} \int n_e^2 dl \quad (46)$$

(Lang 1974). The 0.1 in the -2.1 comes from the Gaunt factor and is approximate but fairly accurate. Thus the free-free emission from the warm ISM may be approximated as a power-law in frequency, $T_{\text{ff}} \propto \nu^{\beta_{\text{ff}}}$ with $\beta_{\text{ff}} \approx -2.1$ nearly independent of temperature. Now we can estimate the confusion caused by Galactic free-free emission. Unfortunately, as figure 11 demonstrates, in the frequency range under consideration, at no frequency does the free-free emission dominate so that it can be unambiguously mapped.

Multi-frequency measurements allow separation of the Galactic and CMB components based on their different spectra. Microwave measurements of the warm ISM complement recombination-line mapping of the same gas. The ratio of free-free antenna temperature to H$\alpha$ line intensity leads to the temperature dependence of the ISM,

$$\frac{T_{\text{ff}} \ (\mu\text{K})}{I_\alpha \ (\text{R})} \approx 520 \ [1 + \sqrt{\frac{T_e}{10^4 \ \text{K}}} \ ] \quad (47)$$

where the H$\alpha$ intensity is in Rayleighs ($10^6/(4\pi)$ photons cm$^{-2}$ s$^{-1}$ sr$^{-1}$). The Wisconsin Hydrogen Alpha Mapper (WHAM) has been funded to map the northern sky with a 1° beam to 10% accuracy in H$\alpha$ (Reynolds, private communication). In combination with a microwave determination of the free-free emission, the temperature of the ISM can be mapped to ~20% precision. The combined data sets will determine whether the intensity variations are attributable solely to density variations, or whether electron temperature (and hence heating mechanism) are significant factors. The combination will allow a good assessment of the free-free component.

However we can estimate how significant the free-free contribution will be as a foreground to CMB measure-ments and determine that there is still a significant window that allows observations to the $10^{-6}$ level. Though the free-free emission never dominates the sky, it does have a sufficiently different spectral index ($\sim -2.1$) than the other components. Thus ratios of maps taken in the cm-wavelength range can reveal regions of enhanced free-free emission. Preliminary analysis of the GEM (Galactic Emission Mapper, De Amici et al. 1995), which has been mapping a substantial portion of the sky at four frequencies: 0.408, 1.5, 2.3, and 5 GHz, shows that the regions of high free-free emission tend to be concentrated in knots on the Galactic plane. Surprisingly, there are a number of such HII regions away from the Galactic plane but they cover a relatively small fraction of the sky. Any free-free emission covering large areas is quite reduced. We can limit the free-free confusion at frequencies above 50 GHz to less than about $10^{-6}$ for all but the largest angular scales.

The low level of free-free power at small angular scales is reminiscent of that for interstellar dust emission probably for the same reason. It is likely that most of the free-free emission is correlated with the density (goes as $n_e^2$) of hydrogen in the Galaxy as are molecular clouds and interstellar dust. A cross-correlation of the COBE DMR data with the DIRBE three lowest frequency bands (dominated by interstellar dust) shows a significant common spatial structure (Kogut et al 1995). Thus the correlated component will have a similar power spectrum to the dust – decreasing at smaller angular scales. The correlated signal is below the 10 $\mu K$ level at 50 GHz for Galactic latitudes above 30°. Most of the power is on the largest angular scales, e.g. the quadrupole. At higher $\ell$ the free-free power is significantly less. This is consistent with the HEMT observations (Tenerife, South Pole, and Saskatoon). Thus we can still be confident that there is a good window through the Galactic foregrounds.

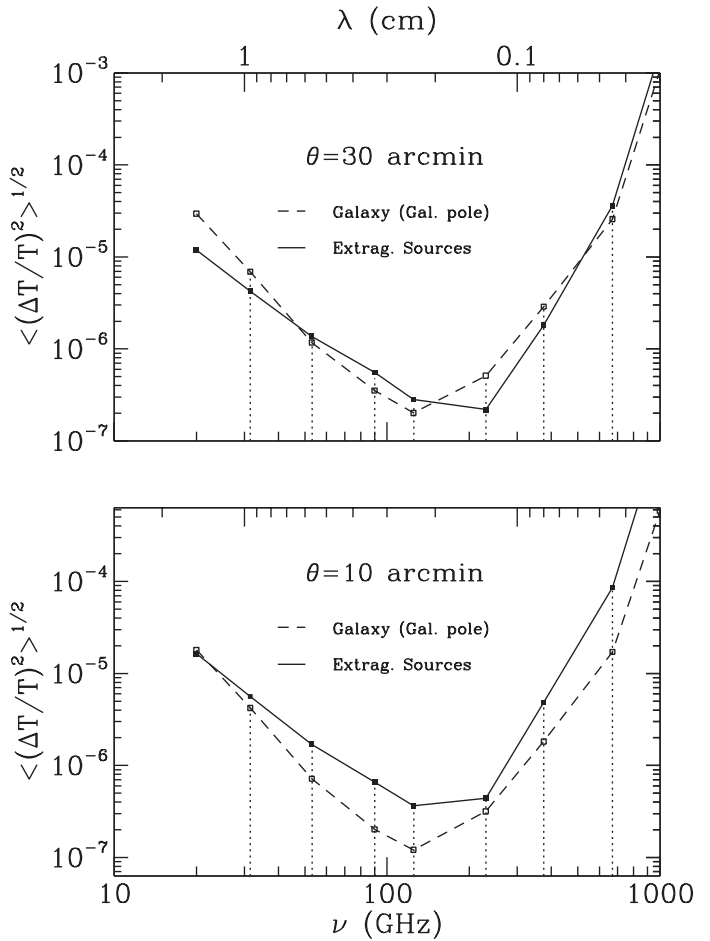

Figure 13: Estimated fluctuation levels due to Galactic polar emission (dashed lines) and to extragalactic sources (solid lines), from Toffolatti et al. (1994). The vertical dotted lines show the COBRAS/SAMBA frequencies.

### Extragalactic Foregrounds

Figure 13 shows the expected fluctuation level at angular resolutions of 10′ and 30′. At high frequency ($\nu \gtrsim 140$ GHz) the main foreground components will be dust emission from infrared cirrus and from normal spiral galaxies, as well as that from starburst galaxies; towards the low-frequency range ($\nu \lesssim 70$ GHz) galactic synchrotron and free–free emission become increasingly important (Toffolatti et al. 1994). Calculations of the residual temperature fluctuations from unresolved extragalactic sources show that in the range 50–300 GHz they will contribute $\Delta T/T \lesssim 10^{-6}$, i.e. below the expected noise level of the detectors. A check (Gawiser & Smoot 1995) based upon the infrared sources from the IRAS catalog shows a high level of consistency with the theoretical predictions of Toffalati et al. Only nearby sources are strong enough to be important and there is a significant frequency range in which the infrared source confusion is below the $10^{-6}$ level. These are readily detected by the higher frequency channels, particularly if the angular resolution increases with frequency.

Extragalactic radio sources are more problematic and a significant number must be excised from the data. Some are known to be variable so that subtraction would require monitoring. This would probably require coordination with ground-based instruments, since these sources are stronger at longer wavelengths where the angular resolution is poorer making them more difficult to detect with the same instrument. Still the fraction of the sky covered by significant sources is small so that for frequencies above 30 to 50 GHz it is likely that for most of the sky the confusion from the sources is below the $10^{-6}$ level.

**Foreground Minimization & Removal**

The Galactic foregrounds may be separated from the CMB by their frequency dependence, supplemented by knowledge of their spatial morphology. This technique is widely used (c.f. Bennett et al. 1992, Brandt et al. 1994) and depends on the frequency coverage and combined uncertainties of the sky surveys used in the fitting process. A limiting factor has been the large calibration uncertainties in the synchrotron-dominated radio surveys (Haslam et al. 1982, Reich and Reich 1986) and the decade gap in frequency coverage between the radio surveys and CMB measurements above 30 GHz (Figure 11). Several authors have pointed out the need for precisely calibrated microwave sky surveys to map the foreground Galactic emission (Bennett et al. 1992, Kogut et al. 1993, Bersanelli et al. 1994, Brandt et al. 1994, Kogut 1995).

**Atmospheric Emission**

Though not a fundamental limitation, as satellites can avoid it, atmospheric emission and especially its variability is an issue for suborbital programs. This is a limitation both for spectrum and anisotropy projects. For ground-based (including mountain-top and Antarctic plateau) experiment design and operational frequency range is limited by atmospheric effects. Clearly some sites are better than others and some techniques allow better elimination of atmospheric effects. However, the atmopshere does drive some experiments to balloon-borne platforms. The effects of the atmosphere and its limitations on CMB experiments are being quantified (Smoot et al. 1987, Church 1994, Bersanelli et al 1995). It is still possible to make significant advances in the suborbital environment. Ultimate scientific goals combined with the limitations imposed by the atmosphere will eventually drive experiments to space-borne platforms.

## §5 FUTURE: 2000+

### 5.1 *The CBR Spectrum – Future*

The analysis and interpretation of present experiments is likely to improve our current knowledge of the spectrum by no more than a factor of three. That still leaves a number of important scientific questions unaddressed.

Answers to this new level of questions require new information on the early universe. It will be necessary to measure the CMB spectrum to $10^{-4}$ precision at centimeter wavelengths. Precise measurements of the CMB spectrum at centimeter wavelengths probe different physical processes than the *COBE* results at millimeter and sub-mm wavelengths and would provide quantitative information on such processes as:

- The transition from an ionized universe to a neutral state and subsequent heating and reionization by the first generation of collapsed objects

- The abundance, lifetimes, and decay modes of hypothesized non-baryonic particles, including supersymmetric partners of known particles or other particle-physics dark-matter candidates

- The decay of primordial turbulence, including the short spatial wavelength end of the observed power spectrum of primordial anisotropy

CMB photons must traverse the interstellar medium within our Galaxy. Precise CMB measurements require an understanding of the diffuse Galactic foregrounds. Figure 11 illustrates the relative intensity of cosmic and Galactic emission at high Galactic latitude. Over a broad range of wavelengths, the CMB dominates the brightness temperature of the sky, and may be distinguished from Galactic sources by its different spectral signature and spatial dependence.

Several factors point to a multi-frequency survey at centimeter wavelengths to 0.01% or better absolute precision as both scientifically interesting and technologically feasible. Measurements in this band probe a poorly surveyed window in the electromagnetic spectrum lying between radio-dish surveys at decimeter wavelengths and *COBE* at millimeter wavelengths. Previous measurements in this band have been limited by interference from the Earth's atmosphere. A multi-channel space experiment with 0.1 mK precision would improve existing data by a factor of 500, provide a decisive test of reionization for redshift $z>50$, produce important constraints to particle dark-matter candidates, and probe physical processes within the Galaxy including the volume partition and heating mechanism of the interstellar medium and the acceleration mechanism of cosmic-ray electrons.

Galactic radio emission is dominated by synchrotron radiation from cosmic-ray electrons and by electron-ion bremsstrahlung (free-free emission) from the ionized interstellar medium (ISM). Despite surveys carried out over many years, relatively little is known about the physical conditions responsible for these diffuse emissions. Synchrotron emission results from cosmic-ray electrons accelerated in magnetic fields, and thus depends on both the electron energy spectrum and the Galactic magnetic field. Radio surveys, at frequencies below a few GHz where synchrotron emission dominates the diffuse sky brightness, map the synchrotron intensity to limited precision but are unable to separate variations in column density, magnetic field, and energy spectrum. Diffuse free-free emission from the ionized ISM *never* dominates the radio sky and is largely unconstrained by observational results. Radio surveys only detect the quadrupolar component and provide almost no information on the spatial power spectrum. Recombination line (H$\alpha$) maps of the same gas suffer from undersampling at high latitudes and self-absorption at low latitudes. A secondary goal of a diffuse emission measurement is to map the spatial dependence of the diffuse Galactic foregrounds, providing answers to outstanding questions on physical conditions in the ISM:

- What is the heating mechanism in the ISM? Is the gas heated by photoionization from the stellar disk, shocks, Galactic fountain flows, or exotic processes such as decaying halo dark matter?

- How is energy injected by supernovae processed? Does the hot gas exist as isolated cavities or chimneys within a pervasive warm background?

- How are cosmic rays accelerated? Is the energy spectrum of local cosmic-ray electrons representative of the Galaxy as a whole?

Achieving 0.01% or better precision across a decade of wavelength requires a design in which instrumental artifacts are minimized or eliminated, imposing several constraints on possible designs. First, it must have an external calibration to compare the power received from the sky with that from an on-board blackbody target, cancelling any instrumental signal to first order. It must have multiple narrow-band channels: separation of a potentially distorted CMB from the Galactic foregrounds based on spectral shape requires at least as many frequency channels as free parameters in any spectral fit. The "interesting" parameters test deviations from a blackbody spectral shape, but are insensitive to the precise value of the undistorted temperature. To this end, the various channels must all observe a common external target: the multiply differential comparison of sky–target between channels is then sensitive only to the target emissivity and temperature gradients but not its absolute temperature. The instrument should be isothermal with the diffuse sky temperature to reduce the effects of small reflection and attenuation within the instrument itself. Finally, it must observe above the bulk of the atmosphere, requiring either a balloon or space platform.

It is clear that for the cm wavelengths that a precision measurement must be made from balloons or space. Attempts from the ground, e.g. GEM, suffer from atmospheric variability in the cm-wavelengths and lack of funding. One proposed instrument to make this measurement is the DIMES (Diffuse Microwave Emission Survey) instrument. DIMES proposes to measure the diffuse cosmic and Galactic emission to 0.1 mK precision in several narrow bands ($\Delta\nu/\nu \sim 10\%$) spanning the poorly surveyed 1–10 cm wavelength range, improving current measurements by a factor of 500. At each frequency, a cryogenic radiometer switched for gain stability between an internal reference load and a beam-defining antenna will measure the signal change as the antenna alternately views the sky and an external blackbody calibration target. By rapidly comparing each channel to the same external target, uncertainties in the absolute target temperature cancel in the derived sky spectra, so that deviations from a blackbody spectral *shape* may be determined much more precisely than the absolute temperature.

The DIMES instrument propose more than two orders of magnitude improvement over the best previous absolute cm-wavelength measurements. Much of this improvement can be attributed to the fully cryogenic design and multiple levels of differences, which cancel instrument gain variations, reflection, and emission to first order. The limiting factor is likely to be the thermal stability of the instrument and the external calibration target: temperature gradients within the target or changes in absolute temperature as the target covers each antenna in turn can mimic spectral and spatial structure in the sky.

## 5.2 The CBR Anisotropy – Future

Progress in the field of cosmology has been extraordinary in the last decade, both in terms of impressive new observations and of consolidation of the theoretical framework of their interpretation. Accurate observations of the cosmic background radiation have played a central role in this progress. In 1992 the COBE team announced the detection of intrinsic temperature fluctuations in the CBR at angular scales larger than $\sim 7°$, with brightness amplitude $\Delta T/T \sim 10^{-5}$ (Smoot et al. 1992).

As outlined above we can expect very significant and rapid progress in the observation of CMB anisotropies. One can expect that the power spectrum of anisotropies will be measured to a sensitivity of about 10% around the first "Doppler" peak, to within a factor of two the cosmic variance at small $\ell$ as a result of various experiments now in the works. Figure 14 shows a sample of the progress one can expect in determining a sample power spectrum in the next few years. To overcome the limit set by sampling variance, it will be necessary to coadd data from multiple balloon flights or observation seasons, since each flight or observing season is likely to cover only 10 to 20% of the sky. Historically this has proven difficult to coadd in this manner as seams appear and calibration may vary. The major advantage of a good satellite mission is the ability to cover the full sky in a single flight or season. The most significant problems in reconstructing the power spectrum will be for those portions where the wavelength is greater than or approximately as large as the surveyed patches. If the CMB anisotropies are from a random-phase stochastic process as assumed when utilizing the power spectrum, then coadding the patches can be expected to give nearly as good a result for the higher $\ell$'s as a single full sky survey. In addition we have a full sky survey from the COBE DMR that gives us a nearly cosmic variance limited power spectrum for $\ell \lesssim 20$. If RELICT 2 is successful, we will have a cross check of the low $\ell$ portion of the power spectrum. Thus we expect these satellite results and the balloon-borne and ground-based (e.g. interferometers) experiments to provide us a good (nearly cosmic variance limited) power spectrum of anisotropies.

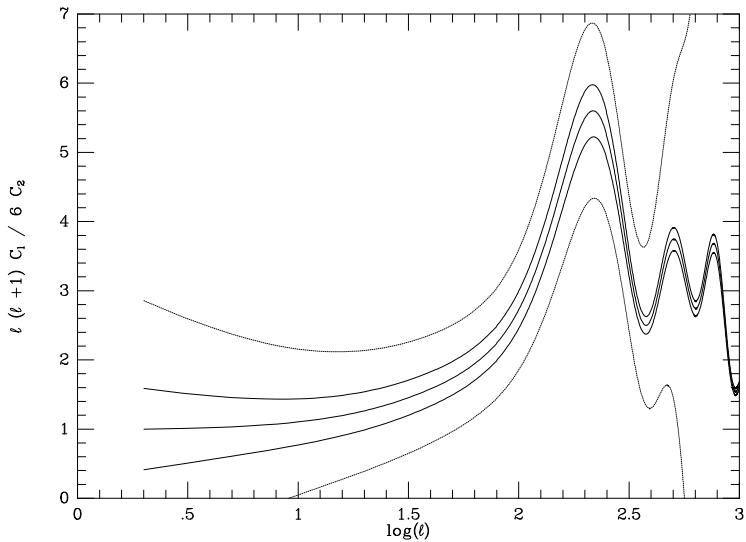

Figure 14: Center line shows the power spectrum for standard CDM ($h = 0.5, \Omega_b = 0.05$, courtsey of Sugiyama, 1995) The pair of lines on either side show the effect of cosmic variance alone. The dotted pair of lines shows the anticipated combined effect of cosmic variance and instrument noise for a single 10-hour flight of the MAX/MAXIMA experiment covering about 10% of the sky. More flights of MAX/MAXIMA, BOOMERANG, and/or ACE should bring down the errors by as much as a factor of 3. The results will come fairly close to cosmic variance limits.

Such results will achieve the first level of scientific goals. The questions are: What science is left? How important is it? And can it be obtained with any approach now forseen and with existing technologies? The short answer is that there is very significant and vital science left to be explored provided the anisotropies and power spectrum can be measured quite accurately. There is a rich theoretical literature exploring various aspects of the issue. The remaining science puts a very high premium on full sky coverage, high sensitivity, and low systematics and residual foregrounds in the data.

A new mission can be expected to have as a product two things: (1) a power spectrum covering the range from about $10'$ to the full sky ($\ell = 2$) which will check and consolidate the previous results with improved systematics and foreground removal. (2) A map of actual anisotropies with good signal to noise. This requires a sensitivity per pixel at about the $\delta T/T \leq 3 \times 10^{-6}$ level (10 $\mu$K) after removal of foregrounds and instrument signature. This is likely to require raw sensitivity about a factor of 3 better than the ultimate quality of the maps and multiple frequency observations spanning the full frequency window and edges for foreground identification.

These are issues that require a space-based platform. There are now four groups actively working on proposed satellite missions to map CMB anisotropies. At the moment three are actively competing directly with each other for a NASA opportunity and much information is kept confidential until the selection. Thus it is instructive to consider the COBRAS/SAMBA mission to understand the scope of a satellite mission, for comparison with the others as they are evaluated, and most importantly to compare with the scientific goals.

COBRAS/SAMBA is currently undergoing Phase A study by ESA. The COBRAS/SAMBA mission is designed for extensive, accurate mapping of the anisotropy of the Cosmic Background Radiation, with angular sensitivity from sub-degree ($\sim 10' - 30'$) scales up to the full sky thus overlapping with the COBE–DMR maps and with signal sensitivity approaching $\Delta T/T \sim 10^{-6}$. This will allow a full identification of the primordial density perturbations which grew to form the large–scale structures observed in the present universe. The COBRAS/SAMBA maps will provide decisive answers to several major open questions relevant to the structure formation epoch and will provide powerful tests for the inflationary model as well as several astrophysical issues. COBRAS/SAMBA will utilize a combination of bolometric and radiometric detection techniques to ensure the sensitivity and wide spectral coverage required for accurate foreground discrimination. An orbit far from Earth has been selected to minimize the unwanted emission from the Earth as a source of contamination.

The COBRAS/SAMBA mission is the result of the merging of two proposals presented in 1993 to the European Space Agency *M3 Call for Mission Ideas*: COBRAS (Cosmic Background Radiation Anisotropy Satellite; Mandolesi et al. 1993) and SAMBA (Satellite for Measurements of Background Anisotropies; Puget et al. 1993). The COBRAS/SAMBA team completed the ESA assessment study in May 1994, and the project continued and is currently in the Phase A study within the European Space Agency M3 programme.

**COBRAS/SAMBA Scientific Objectives**

The COBRAS/SAMBA mission will produce near all-sky maps of the background anisotropies in 8 frequency bands in the range 30–800 GHz, with peak sensitivity $\Delta T/T \sim 10^{-6}$. The maps will provide a detailed description of the background radiation fluctuations. Individual hot and cold regions should be identified above the statistical noise level, at all angular scales from $\lesssim 10'$ up to very large scales, thus providing a high resolution imaging of the last scattering surface.

The COBRAS/SAMBA maps will provide all multipoles of the temperature anisotropies from $\ell = 1$ (dipole term) up to $\ell \simeq 1500$ (corresponding to $\sim 7'$). It is the information contained in this large number of multipoles that can probe the various proposed scenarios of structure formation and the shape of the primordial fluctuation spectrum (for comparison, the COBE–DMR maps are limited to $\ell \lesssim 20$).

The high resolution COBRAS/SAMBA maps will provide a key test for structure formation mechanisms, based on the statistics of the observed $\Delta T/T$ distribution. The inflationary model predicts Gaussian fluctuations for the statistics of the CBR anisotropies, while alternative models based on the presence of topological defects, such as strings, monopoles, and textures, predict non–Gaussian statistics (e.g. Coulson et al. 1994). Due to the different nature of their early history causality constrains primordial perturbations from a source such as inflation and from topological defects to have a different anisotropy power spectra particularly in the region of the "Doppler" peaks (Albrecht et al. 1995). The angular resolution and sensitivity of COBRAS/SAMBA will allow discrimination between these alternatives with tests of both the power spectrum and statistics.

The high-order multipoles will allow an accurate measure of the spectral index $n$ of the primordial fluctuation spectrum:

$$(\delta\phi)^2 \propto \lambda^{(1-n)} \qquad (48)$$

where $\delta\phi$ is the potential fluctuation responsible for the CBR anisotropies, and $\lambda$ is the scale of the density perturbation. This corresponds to CBR temperature fluctuations $(\Delta T/T)^2 \propto \theta^{(1-n)}$ for angles $\theta > 30' \, \Omega_0^{1/2}$. The proposed observations will be able to verify accurately the scale invariant "Harrison–Zel'dovich" spectrum ($n = 1$) predicted by inflation. Any significant deviation from that value would have extremely important consequences for the inflationary paradigm. The COBE–DMR limit on the spectral index after two years of observations ($n = 1.1^{+0.3}_{-0.4}$, 68% CL; Gorski et al. 1994) can be constrained $\sim 10$ times better by the COBRAS/SAMBA results.

The proposed observations will provide an additional, independent test for the inflationary model. Temperature anisotropies on large angular scales can be generated by gravitational waves (tensor modes, $T$), in addition to the energy-density perturbation component (scalar modes, $S$). Most inflationary models predict a well determined, simple relation between the ratio of these two components, $T/S$, and the spectral index $n$ (Davis et al. 1992, Little & Lyth 1992):

$$n \approx 1 - \frac{1}{7}\frac{T}{S}. \qquad (49)$$

The COBRAS/SAMBA maps will be able to verify this relationship, since the temperature anisotropies from scalar and tensor modes vary with multipoles in different ways.

A good satellite mission will be able not only to test the inflationary concept but also to distinguish between various models and determine inflationary parameters. There is an extensive literature on what can be determined about inflation such as the scalar and tensor power spectra, the energy scale of inflation and so on (see e.g. Steinhardt 1995, Knox 1995). Such quality measurements lead also to good observations or constraints for $\Omega_0$, $\Omega_{baryon}$, $\Lambda$, $H_0$, etc. Sub–degree anisotropies are sensitive to the ionization history of the universe. In fact, they can be erased if the intergalactic medium underwent reionization at high redshifts. Moreover, the temperature anisotropies at small angular scales depend on other key cosmological parameters, such as the initial spectrum of irregularities, the baryon density of the universe, the nature of dark matter, and the geometry of the universe (see e.g. Crittenden et al. 1993, Bond et al. 1993, Kamionkowski et al. 1994 Hu & Sugiyama 1994, Scott, Silk, & White 1995)). The COBRAS/SAMBA maps will provide constraints on these parameters within the context of specific theoretical models.

Moreover, COBRAS/SAMBA should measure the Sunyaev–Zel'dovich effect for more than 1000 rich clusters, using the higher resolution bolometric channels. Combined with X–ray observations these measurements can be used to estimate the Hubble constant $H_0$ as a second independent determination.

**Foreground Emissions**

In order to obtain these scientific goals, the measured temperature fluctuations need to be well understood in terms of the various components that add to the cosmological signal. In fact, in addition to the CBR temperature fluctuations, foreground structures will be present from weak, unresolved extragalactic sources and from radiation of galactic origin (interstellar dust, free–free and synchrotron radiation).The COBRAS/SAMBA observations will reach the required control on the foreground components in two ways. First, the large sky coverage ($\geq 90\%$ of the sky) will allow accurate modeling of these components where they are dominant (e.g. galactic radiation near the galactic plane). Second, the observations will be performed in a spectral range as broad as possible.

In fact, the COBRAS/SAMBA channels will span the spectral region of minimum foreground intensity (in the range 50–300 GHz), but with enough margin at high and low frequency to monitor "in real–time" the effect of the various foreground components (see e.g. Brandt et al. 1994). By using the COBRAS/SAMBA spectral information and modeling the spectral dependence of galactic

| Table COBRAS/SAMBA Payload Characteristics | | | | | | | | |
|---|---|---|---|---|---|---|---|---|
| Telescope | 1.5 m Diam. Gregorian; system emissivity ≤1% | | | | | | | |
| | Viewing direction offset ≥ 70° from spin axis | | | | | | | |
| Instrument | LFI | | | | HFI | | | |
| Center Frequency (GHz) | 31.5 | 53 | 90 | 125 | 140 | 222 | 400 | 714 |
| Wavelength (mm) | 9.5 | 5.7 | 3.3 | 2.4 | 2.1 | 1.4 | 0.75 | 0.42 |
| Bandwidth ($\frac{\Delta \nu}{\nu}$) | 0.15 | 0.15 | 0.15 | 0.15 | 0.4 | 0.5 | 0.7 | 0.6 |
| Detector Technology | HEMT receiver arrays | | | | Bolometers arrays | | | |
| Detector Temperature | ~ 100 K | | | | 0.1 - 0.15 K | | | |
| Cooling Requirements | Passive | | | | Cryocooler + Dilution system | | | |
| Number of Detectors | 13 | 13 | 13 | 13 | 8 | 11 | 16 | 16 |
| Angular Resolution (arcmin) | 30 | 20 | 15 | 12 | 10.5 | 7.5 | 4.5 | 3 |
| Optical Efficiency | 1 | 1 | 1 | 1 | 0.3 | 0.3 | 0.3 | 0.3 |
| $\frac{\Delta T}{T}$ Sensitivity ($1\sigma, 10^{-6}$ units, 90% sky coverage, 2 years) | 1.7 | 2.7 | 4.1 | 7.2 | 0.9 | 1.0 | 8.2 | $10^4$ |
| $\frac{\Delta T}{T}$ Sensitivity ($1\sigma, 10^{-6}$ units, 2 % sky coverage, 2 years) | 0.6 | 0.9 | 1.4 | 2.4 | 0.3 | 0.3 | 2.7 | 5000 |

Table 1: Instrumental Parameters for COBRAS/SAMBA the most important factors are the frequency coverage, the angular resolution, sky coverage, and sensitivity.

Assuming:
$\theta = 0.5°$
array with N=10
$\delta T = 0.5$ mK Hz$^{-1/2}$

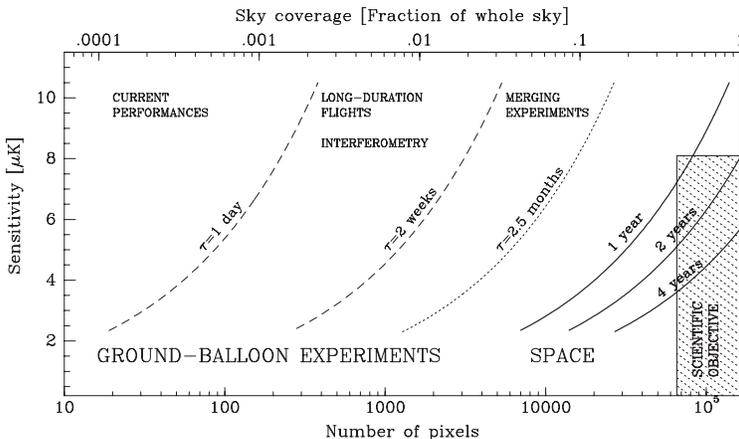

Figure 15: Number of pixels covered versus sensitivity for various potential platforms.

and extragalactic emissions it will be possible to remove the foreground contributions with high accuracy.

It should be noted that in most channels the final limitation to the cosmological information of the COBRAS/SAMBA maps is expected to be due to the residual uncertainties in the separation of the foreground components rather than statistical noise. This explains why the overall design of the instrument and payload is highly driven by the need of achieving a spectral coverage as large as possible. Performing measurements where the dominant foreground components are different will permit a powerful cross check on residual systematic errors in the CBR temperature fluctuation maps.

The need of accurate characterization of all non-cosmological components, of course, brings the benefit of additional astrophysical information. The very large COBRAS/SAMBA data base, particularly when combined with the IRAS survey, can provide information on several non-cosmological issues, such as the evolution of starburst galaxies, the distribution of a cold-dust component, or the study of low-mass star formation.

### The Payload

The COBRAS/SAMBA model payload consists mainly of a shielded, off-axis Gregorian telescope, with a parabolic primary reflector and a secondary mirror, leading to an integrated instrument focal plane assembly. The payload is part of a spinning spacecraft, with a spin rate of 1 rpm. The focal plane assembly is divided into low-frequency (LFI) and high-frequency (HFI) instrumentation according to the technology of the detectors. Both the LFI and the HFI are designed to produce high-sensitivity, multi-frequency measurements of the diffuse sky radiation. The LFI will measure in four bands in the frequency range 30–130 GHz (2.3–10 mm wavelength). The HFI will measure in four channels in the range 140–800 GHz (0.4–2.1 mm wavelength). The highest frequency LFI channel and the lowest HFI channel overlap near the minimum foreground region. Table 1 summarizes the main characteristics of the COBRAS/SAMBA payload.

## The Main Optical System

A clear field of view is necessary for the optics of a high-sensitivity CBR anisotropy experiment to avoid spurious signals arising from the mirrors or from supports and mechanical mounting. A Gregorian configuration has been chosen, with an primary parabolic mirror of 1.5 meter, and an elliptic secondary mirror (0.57 m diameter). Stray satellite radiation and other off–axis emissions are minimized by underilluminating the low–emissivity optics. The telescope reimages the sky onto the focal plane instrument located near the payload platform. The telescope optical axis is offset by 70° or more from the spin axis. Thus at each spacecraft spin rotation the telescope pointing direction sweeps a large (approaching a great) circle in the sky, according to the sky scan strategy.

Blockage is a particularly important factor since several feeds and detectors are located in the focal plane, and unwanted, local radiation (e.g. from the Earth, the Sun and the Moon) needs to be efficiently rejected. A large, flared shield sorrounds the entire telescope and focal plane assembly, to screen the detectors from contaminating sources of radiation. The shield also plays an important role as an element of the passive thermal control of the spacecraft.

## The Focal Plane Assembly

The necessary wide spectral range requires the use of two different technologies, bolometers and coherent receivers incorporated in a single instrument. Both technologies have shown impressive progress in the last ten years or so, and more is expected in the near future. The thermal requirements of the two types of detectors are widely different. The coherent radiometers (LFI), operating in the low frequency channels, give good performance at operational temperature of $\sim$ 100 K, which is achievable with passive cooling. The bolometers, on the other hand, require temperatures $\leq$ 0.15 K in order to reach their extraordinary sensitivity performances. The main characteristics of the LFI and HFI are summarized in Table 1.

The LFI consists of an array of 26 corrugated, conical horns, each exploited in the two orthogonal polarization modes, feeding a set of state–of–the–art, high sensitivity receivers. The receivers will be based on MMIC (Monolithic Microwave Integrated Circuits) technology with HEMT (High Electron Mobility Transistor) ultra–low noise amplifiers (see e.g. Pospieszalski et al. 1993). Since the whole LFI system will be passively cooled, it can be operated for a duration limited only by spacecraft consumables (up to 5 years). The three lowest center frequencies of the LFI were chosen to match the COBE-DMR channels, to facilitate the comparison of the product maps.

About 50 bolometers will be used in the HFI instrument, which require cooling at $\sim$ 0.1 K. The cooling system combines active coolers reaching 4 K with a dilution refrigeration system working at zero gravity. The refrigeration system will include two pressurized tanks of $^3$He and $^4$He for an operational lifetime of 2 years.

## Orbit and Sky Observation Strategy

One of the main requirements for the COBRAS/SAMBA mission is the need of a far–Earth orbit. This choice greatly reduces the problem of unwanted radiation from the Earth which is a serious potential contaminant at the high goal sensitivity and angular resolution. The requirements on residual Earth radiation are basically the same for the LFI and the HFI systems. Adopting a low–earth orbit, such as that used by the COBE satellite, the requirement on straylight and sidelobe rejection would be a factor of $10^{13}$, which is beyond the capabilities of present microwave and sub–mm systems and test equipment. Two orbits have been considered for COBRAS/SAMBA: a small orbit around the L5 Lagrangian point of the Earth–Moon system, at a distance of about 400,000 km from both the Earth and the Moon and the L2 Lagrange point of the Earth–Sun system. From the Earth–Moon Lagrange point the required rejection is relaxed by four orders of magnitude, which is achievable with careful, standard optical designs. For the Earth–Sun L2 point the situation for the Earth and Moon is even better and the Sun is basically unchanged but because the Earth, Moon, and Sun are all roughly in the same direction, the spacecraft can be oriented very favoably.

These orbits are also very favorable from the point of view of passive cooling and thermal stability (Farquhar & Dunham 1990). The spacecraft will be normally operated in the anti-solar direction, with part of the sky observations performed within ±40° from anti–solar.

Other potential missions considered both a heliocentric orbit and the Earth–Sun L2 point. All concerned seemed to have come to the conclusion that the Earth–Sun L2 point is the best choice. Operationally, it is difficult to find a more optimum location.

The main goal of the mission is to observe nearly the whole sky ($\gtrsim$ 90%) with a sensitivity of 10–15 $\mu$K within the two year mission lifetime. Deeper observation of a limited ($\sim$ 2%) sky region with low foregrounds could significantly contribute to the cosmological information. Simulations have shown that these observational objectives can be achieved simultaneously in a natural way, using the spinning and orbit motion of the spacecraft, with relatively simple schemes.

## §6 Interpretation, Future

In three short years the field of CMB anisotropy observations and theory has made great strides. Until April 1992 all plots of CMB anisotropy showed only upper limits, except for the $\ell = 1$ dipole. Now we are beginning to trace out the shape of the power spectrum and to make maps of the anisotropies. This promises to deliver a wealth of new information to cosmology and to connect to other fields. The COBE DMR has now released the first two years of its data and the full four-year data set is being processed and prepared for release in fall 1995. We can expect improved results from the DMR on the large angular scales but the scientific interest has moved to covering the full spectrum and learning what the medium and small angular scales will tell us. Already we are seeing plots showing the CMB anisotropy spectrum related to and overlaid on the primordial density perturbation power spectrum and attempts to reconstruct the inflaton potential. These are the first steps in a new period of growth.

Experiments are underway. Nearly every group has data under analysis and is also at work on developing new experiments. The first of these are the natural extensions of the ongoing experiments. Some groups are considering novel approaches. Real long-term progress depends on avoiding the potential foregrounds: fluctuations of the atmosphere, a source of noise that largely overwhelms recent advances in detector technology, and Galactic and extragalactic signals. This requires instruments having sufficient information (usually only through multifrequency observations) and observing frequencies to separate out the various components. It also means going above the varying atmosphere. Collaborations are working on long-duration ballooning instruments. Ultimately, as COBE has shown, going to space really allows one to overcome the atmospheric problem and to get data in a very stable and shielded environment. A number of groups are working on designs for new satellite experiments. The COBRAS/SAMBA mission (Mandolesi et al. 1994) leads the way in the multi-wavelength and benign orbit location. With the new data that are appearing, can be expected, and ultimately will come from the COBRAS/SAMBA mission we can look forward to a very significant improvement in our knowledge of cosmology.

An accurate, extensive imaging of CBR anisotropies with sub–degree angular resolution would provide decisive answers to several major open questions on structure formation and cosmological scenarios. The observational requirements of such an ambitious objective can be met by a space mission with a far–Earth orbit and instruments based on state–of–the–art technologies.

Atmospheric disturbance, emission from the Earth and limited integration time are the main limiting factors which prevent ground–based and balloon–borne experiments from obtaining sufficient sensitivity over very large sky regions, with additional difficulties in reaching accurate foreground removal (see Danese et al. 1995 for a recent discussion). Only a suitably designed space mission can meet the scientific goals outlined in section 2. On the other hand it should be stressed that experiments from the ground or from balloons are not alternative to a space mission like COBRAS/SAMBA, but rather complementary.

## §7 Recommendations

A strong vigorous program of CMB observations should be supported. The field, especially CMB anisotropies, is very active and fertile at the present and stands at the threshhold of results that will revolutionize cosmology and point the way to future. We can anticipate that critical new observations will result in breakthroughs in our understanding by about the year 2000.

There will remain more high-value science that is best approached by CMB observations. To make a quantum step forward will require space-based missions. Both the CMB spectrum and anisotropy are open for major advances. The polarization of the CMB is likely to move forward as a piggy-back effort on high-quality anisotropy experiments. The area theory currently shows to be very rich is the detailed study of CMB anisotropies.

To make the appropriate quantum step over what is achievable by existing and propose CMB anisotropy instuments a satellite mission has to excel in a number of areas:

- full sky coverage $\geq 90\%$
- high sensitivity - $\leq 10\mu K$ per pixel
- good angular resolution - $\lesssim 10'$
- low residual foregrounds
- low systematics.

These features will be necessary as one can anticipate that suborbital programs will map out the CMB anisotropy power spectrum to nearly the cosmic variance limit. What the satellite project aims to do is improve the quality of the data and actually map the anisotropies with reasonable signal-to-noise ratios. Thus high quality results in terms of low residuals and low systematics given by uniform sky coverage with a single wide-spectral range, high sensitivity instrument in a single well-calibrated mission is the key issue. This will provide confidence and a check of the results to that point and will provide a map of anisotropies for statistical and morphological study. It should represent a major consolidation of our knowledge of CMB anisotropy and cosmology.

# Acknowledgments

Thanks to Douglas Scott and Martin White for providing the plot of the current state of CMB observations and comments, Marco Bersanelli and Al Kogut for wording and plots and Laura Cayon for assistance in providing plots of the results including cosmic variance and instrument noise. This work is supported in part by the Director, Office of Energy Research, Office of High Energy and Nuclear Physics, Division of High Energy Physics of the U.S. Department of Energy under Contract No. DE-AC03-76SF00098.